\documentclass[a4paper,11pt]{article}
\usepackage[hypcap=false]{caption}

\usepackage{jheppub}     
\usepackage[T1]{fontenc} 
\usepackage{bm}
\usepackage{mathrsfs}
\usepackage{cancel}
\usepackage[normalem]{ulem}
\usepackage{slashed, braket, dirtytalk, mathtools}
\usepackage{xcolor}
\usepackage{float, soul, parskip, lmodern}
\usepackage{caption}
\usepackage{siunitx}
\usepackage{comment}
\usepackage{subcaption}
\usepackage{natbib}
\usepackage{bbold}
\usepackage{tikz}
\usepackage[compat=1.1.0]{tikz-feynman}
\usepackage{booktabs}
\hypersetup{colorlinks=true}

\relax

\def\be{\begin{equation}}
\def\ee{\end{equation}}

\def\sla{\raise.15ex\hbox{$/$}\kern-.57em}

\newcommand{\bear}{\begin{eqnarray}}
\newcommand{\bea}{\begin{eqnarray}}
\newcommand{\eea}{\end{eqnarray}}
\newcommand{\bi}{\begin{itemize}}
\newcommand{\ei}{\end{itemize}}
\newcommand{\ba}{\begin{array}}
\newcommand{\ea}{\end{array}}
\newcommand{\bcomm}{}

\newbox\pippobox

						\def\d{\delta}
\def\e{\varepsilon}	\def\z{\zeta}					
						\def\m{\mu}
\def\n{\nu}						\def\p{\pi}			\def\r{\rho}	
\def\s{\sigma}								 

\def\ie{{\it i.e.}}
\def\eg{{\it e.g.}}
\def\<{\big\langle}
\def\>{\big\rangle}

\def\simlt{\mathrel{\lower2.5pt\vbox{\lineskip=0pt\baselineskip=0pt
           \hbox{$<$}\hbox{$\sim$}}}}
\def\simgt{\mathrel{\lower2.5pt\vbox{\lineskip=0pt\baselineskip=0pt
           \hbox{$>$}\hbox{$\sim$}}}}

\def\nn{\nonumber}

\newcommand{\esp}{\phantom{\!\!\overset{\displaystyle |}{|}}}
\newcommand{\dis}{\displaystyle}

\def\simlt{\mathrel{\lower2.5pt\vbox{\lineskip=0pt\baselineskip=0pt
           \hbox{$<$}\hbox{$\sim$}}}}
\def\simgt{\mathrel{\lower2.5pt\vbox{\lineskip=0pt\baselineskip=0pt
           \hbox{$>$}\hbox{$\sim$}}}}

\title{\boldmath Anomalous $U(1)$ extension of the Standard Model}

\preprint{}

\author[a]{Pascal Anastasopoulos,}
\author[b,c]{Ignatios Antoniadis,}
\author[c]{Karim Benakli}
\author[d]{and Fran\c{c}ois Rondeau}

\affiliation[a] {Current affiliation: European Research Council Executive Agency (ERCEA),\\ 
Place Charles Rogier 16/1 1210 Brussels
Belgium\\
Disclaimer: The views expressed are purely those of the authors and may not in any circumstances be regarded as stating an official position of the ERCEA and
the European Commission.\\
\& \\
Institute of High Energy Physics, 
  Austrian Academy of Sciences, \\ Georg - Coch - Platz 2, 
  1010 Vienna, Austria.}

\affiliation[b] {High Energy Physics Research Unit, Faculty of Science, Chulalongkorn University, Bangkok 1030, Thailand.}

\affiliation[c] {Sorbonne Universit\'e, CNRS, Laboratoire de Physique Th\'eorique et Hautes Energies, LPTHE, F-75005 Paris, France.}

\affiliation[d]{Department of Physics, University of Cyprus,\\ Nicosia 1678, Cyprus.}

\emailAdd{paschalis.anastasopoulos@oeaw.ac.at}
\emailAdd{antoniad@lpthe.jussieu.fr}
\emailAdd{kbenakli@lpthe.jussieu.fr}
\emailAdd{rondeau.francois@ucy.ac.cy}


\abstract{We present a set of example models in which the Standard Model (SM) symmetry group is extended by a new abelian symmetry. This additional symmetry appears anomalous in the effective low-energy theory; however, the anomalies cancel out when massive chiral fermions not present in the effective low-energy theory are taken into account. These chiral fermions under the new abelian gauge group, are chosen to be vector-like under the SM symmetries, and reside in the same representations as quarks and leptons.  This allows us to quantitatively determine 
the magnitude of tree-level interactions between three vector bosons induced in low-energy effective field theory by the integration of chiral heavy fermions. We also examine the perturbativity constraints of the theory and the ultraviolet cut-off. We conclude by highlighting possible extensions of our work. }

\begin{document}
\maketitle
\flushbottom

\newpage
\section{Introduction}

Gauge theories provide a framework for describing the strong, weak, and electromagnetic interactions of the Standard Model (SM) of particle physics in a manifestly Lorentz-invariant way. While gauge symmetries are initially imposed at the classical level, their survival upon quantization is not assured due to the problem of gauge anomalies. Gauge symmetries are necessary to eliminate unphysical longitudinal modes of the gauge bosons. Anomalies would compromise the consistency of the entire theory unless these modes are made physical. This occurs when the symmetries are spontaneously broken, rendering the gauge bosons massive. Concepts related to these situations are extensively covered in existing literature; interested readers can refer, for instance, to \cite{DHoker:1984izu,DHoker:1984mif,GREEN1984117,ALVAREZGAUME1984269,Preskill:1990fr,Batra:2005rh,Kumar:2007zza,Antoniadis:2009ze,Dedes:2012me,Dror:2017ehi,Dror:2017nsg,Allanach:2018vjg,Costa:2019zzy,Costa:2020dph,Costa:2020krs,Allanach:2020zna,Allanach:2019uuu,Allanach:2019gwp,Alvarez-Gaume:2022aak}.

In this work, we focus on an extension of the SM by a massive gauge boson denoted as $Z^{\prime}$. The potential discovery of this additional boson may not necessarily coincide with the detection of all corresponding fermions charged under the associated abelian gauge group. At experimentally achievable energy scales, it is possible that only a subset of these fermions might be observable, potentially leading to anomalous outcomes when analyzing contributions to triangular Feynman diagrams. However, upon considering the complete model, encompassing all fermions with masses falling below a fundamental scale of quantum gravity -- such as the string or Planck mass -- the anomalies might not persist. While such a scenario and its  features have been discussed, a concrete realization of an extended SM model exhibiting these characteristics remains absent. Therefore, our objective here is to explicitly construct models that embody such attributes. Phenomenological studies of extra $Z^{\prime}$ gauge bosons can be found in \cite{Cvetic:1995rj,Cvetic:1997mbh,Kiritsis:2002aj,Ghilencea:2002da,Kors:2004dx,Kors:2004ri,Kors:2005uz,Feldman:2006ce,Coriano:2005own,Anastasopoulos:2008jt,Armillis:2008vp,Accomando:2016sge,Dudas:2012pb,Dudas:2013sia,Dudas:2009uq,Ekstedt:2017tbo,Buras:2021btx,Davighi:2021oel,Anchordoqui:2021vrg,Anchordoqui:2021lmm,Antoniadis:2021mqz, Arun:2022ecj,Mandal:2023mck}.

As a concrete example, we will consider a minimal extension of the SM gauge symmetry by one additional abelian factor labeled as $U(1)_A$.  The matter fields consist of the SM sector content alongside a new sector of supplementary fermions in the same $SU(3)$ and $SU(2)$ representations as those in the SM but with distinct hypercharges. These additional fermions appear vector-like under the SM gauge group, ensuring an automatic anomaly cancellation of all reducible and irreducible anomalies of the $SU(3)\times SU(2)\times U(1)_Y$ gauge symmetry. The fermions carry charges under $U(1)_A$ in a manner that the model is anomaly-free, but  that anomaly cancellation does not occur separately to the two sectors, but only after considering the collective contribution of chiral fermions across both sectors. A scalar field $S$ acquires a vacuum expectation value (vev) that breaks $U(1)_A$ symmetry and generates a mass for $Z^{\prime}$.

The evaluation of anomalies involves computing the triangular Feynman diagrams with external legs at the vertices corresponding to three vector bosons $i$, $j$, and $k$. The result of any 
diagram is proportional to a combination of charges that we denote as $t_{ijk}$. As anomaly cancellation does not occur separately within the two sectors, in order for all anomalies to cancel, we need
\bea
\sum_{\textrm{total}} t_{ijk} = \sum_{\textrm{SM}} t_{ijk}+ \sum_{\textrm{extra}} t_{ijk} =0~~~~~ \textrm{with}~~~~~ \sum_{\textrm{SM}} t_{ijk}= - \sum_{\textrm{extra}} t_{ijk}\neq 0, 
\label{anomaly conditions in general}
\eea
where $i,j,k$ correspond now to the gauge factors from $SU(3)\times SU(2)\times U(1)_Y\times U(1)_A$ and the sum refers to all individual internal contributions. 

\begin{figure}[h!]
\includegraphics[height=63mm]{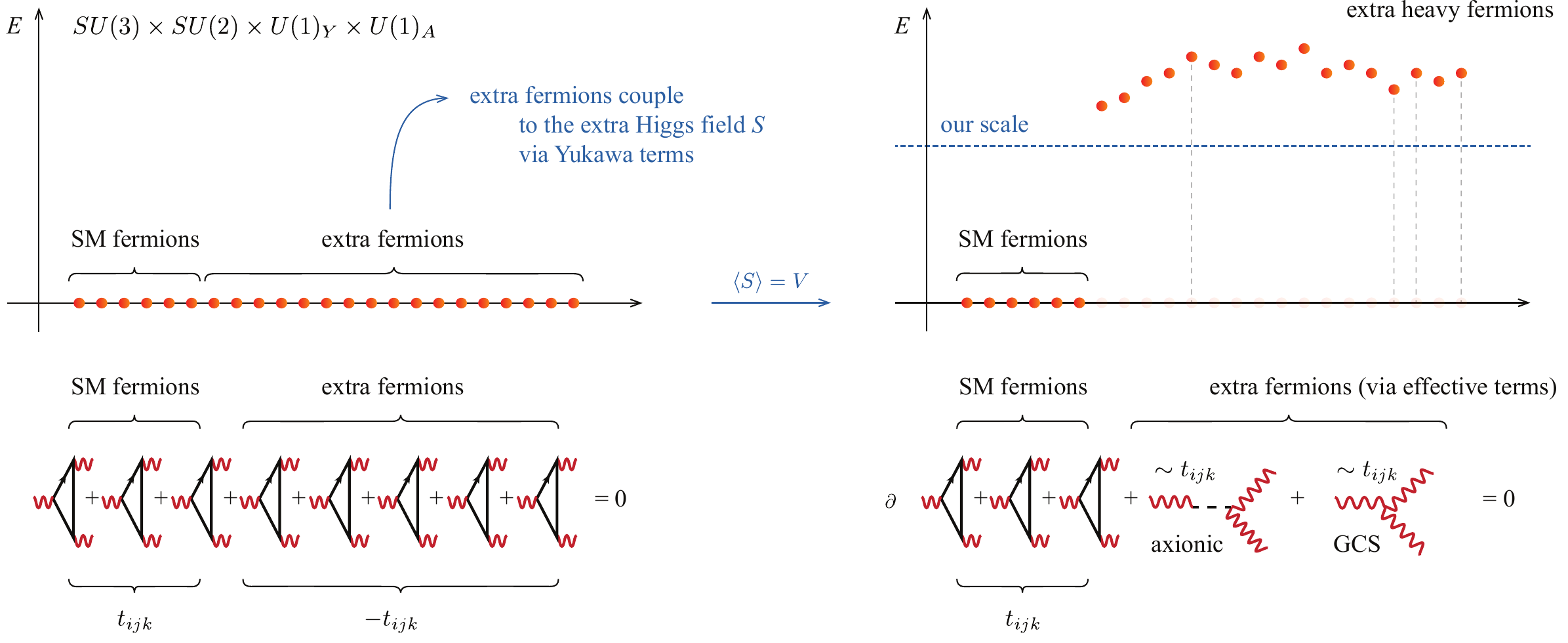}
\caption{Left panel: in the unbroken phase, both the SM and extra fermions are massless. The anomalous contributions from the extra fermions cancel the ones from the SM fermions, so that the model is anomaly free ($t_{ijk}^\textrm{SM}+t_{ijk}^\textrm{extra}=0$). Right panel: after the extra Higgs $S$ gets a vev, the extra fermions become massive. At energies below their mass, the SM might seem anomalous ($t_{ijk}^\textrm{SM}\neq 0$). However, effective axionic and GCS terms contribute in order to cancel this anomaly. In both figures, the vertical axis $E$ depicts the energy scale.} \label{figure:Before And After}
\end{figure}

It is interesting to consider the anomaly cancellation in the two phases of the model: for $\braket{S}=0$ and $\braket{S} \neq 0$. These are illustrated in Figure \ref{figure:Before And After}.
Let's assume that all of the extra fermions have Yukawa couplings of order one with $S$. For $\braket{S} \neq 0$, they become massive. If the vacuum expectation value of $S$ is larger than the energy scales at reach, then the extra fermions will be integrated out, and their corresponding contributions to triangular diagrams will appear in the low-energy effective field theory as effective three-point interactions between vector bosons. Some appear as exchange of axions, while 
others present themselves as generalized Chern-Simons (GCS) terms. Such couplings have been independently discovered in the study of D-brane realizations of the SM \cite{Antoniadis:2000ena,Antoniadis:2001np}, in supergravity \cite{DEWIT1985569,Andrianopoli:2002mf,Andrianopoli:2004sv,deWit:2002vt} as well as in higher dimensional gauge theories \cite{Dudas:2003iq,Dudas:2004ni,Scrucca:2004jn,Scrucca:2001eb,Hill:2006ei}. A detailed analysis of these terms has been carried out both in field theory and string theory in \cite{Anastasopoulos:2006cz}, see also \cite{Anastasopoulos:2007qm} for a review. The sum of all contributions to the three-point functions ensures the preservation of gauge symmetries Ward-Takahashi identities, and the conservation of corresponding currents, as expected from the complete model being anomaly-free. 

\begin{figure}[h!]
\includegraphics[height=25mm]{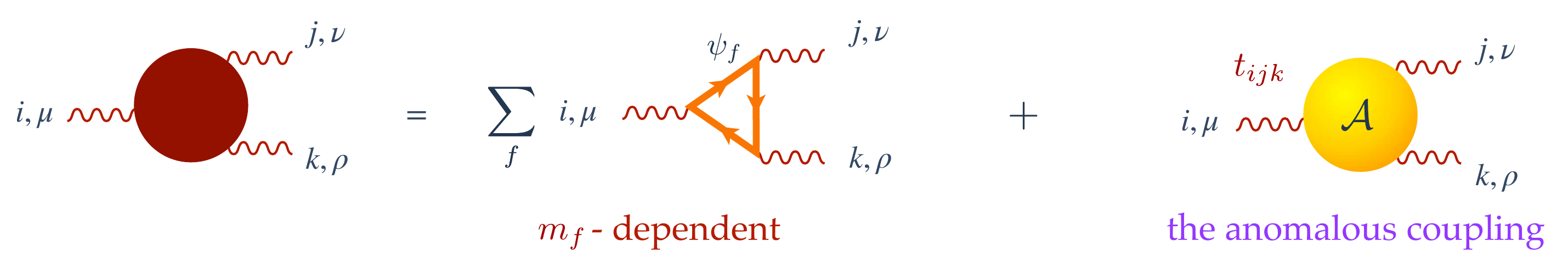}
\caption{Decomposition of the three-point function involving three external gauge bosons as the sum of a set of terms depending on the mass $m_f$ of the fermions, and a mass-independent set of terms containing the anomalous couplings.} \label{fig:feyn_diag}
\end{figure}

The three-point function involving three external gauge bosons can be written as the sum of two sets of terms:
\begin{itemize}
\item One-loop triangular diagrams involving SM fermions running in the loop. These terms depend on the individual charges of the fermions as well as their masses: the heavier the fermion, the smaller the contribution. They are always present, whether the model is anomalous or not.
\item The effective anomalous couplings, containing the axionic and GCS terms, which arise after integrating out the heavy fermions. These terms depend on the anomaly coefficients $t_{ijk}$, and thus vanish in anomaly-free models. They crucially do not depend on any mass.
\end{itemize}
This is depicted on Figure \ref{fig:feyn_diag}.

Therefore, if the anomaly coefficients are large (due to the contribution of several heavy fermions), the axionic and GCS terms become very large. The anomalous coupling then becomes significant and can compete with the $m_f$-dependent terms. This might have interesting phenomenological consequences, considering for instance the decay $Z^{\prime}\rightarrow Z \gamma$.

Large anomalous couplings might also have interesting phenomenological consequences on the anomalous magnetic moment of the muon ($(g-2)_{\mu}$), as described in \cite{Anastasopoulos:2022ywj}. Considering the contributions of an anomalous $Z^{\prime}$ gauge boson to the $(g-2)_{\mu}$, it has been shown that there exists regions of the parameter space where the anomalous (two-loop) contribution becomes comparable and even dominant compared to the one-loop contribution. This situation occurs for ``large'' anomaly coefficients, of order $t_{ijk}\sim\mathcal{O}(10-10^2)$. This motivates that we provide explicit examples with numerical values for the different $t_{ijk}$'s, showing that the setup described in \cite{Anastasopoulos:2022ywj} can be realized in realistic models.


This work is structured as follows. In Section~\ref{sect:fundamental_model}, we introduce the models and explicitly outline the conditions for anomaly cancellation. Section~\ref{sect:anomaly_free_sol} is dedicated to solving these conditions, providing solutions for a set of illustrative examples. Section \ref{sect:effective_model} delves into the discussion of low-energy three-vector bosons effective operators. The ultraviolet (UV) cut-off of the effective theory and the running of couplings are discussed in Section~\ref{cut-off section}. This not only imposes constraints on the number of additional fermions to maintain perturbativity but also offers insights into the hierarchy between gauge and Yukawa couplings, highlighting the advantages of our choice of extra fermions charged under the SM gauge group. The paper concludes with a brief outlook on possible future work and summarizes our findings.

\section{Anomaly-free Models}
\label{sect:fundamental_model}

We explore an extension of the SM group $SU(3)\times SU(2)\times U(1)_Y$ with an extra abelian factor, referred to as $U(1)_A$. The respective gauge couplings are denoted as $g_3,g_2,g_Y$ and $g_A$. The model particle content is partitioned into two distinctive sectors: the SM sector and a ``secluded sector''. The latter consists in an ensemble of supplementary fermions denoted as $\psi$, deliberately adopting the same representations as the SM matter fields.  

We consider a scenario in which the typical energies $E$ within our effective field theory description result in the absence of some or all of the extra fields, as they possess masses $M_\psi \gg E$. Our objective is to orchestrate a situation in which the contributions to the anomalies of the $U(1)_A$ gauge symmetry cancel out between the light fields present in the effective field theory and the (non-observable) heavier chiral fermions. Our intention is not to provide an exhaustive classification of all possible models that can be derived from the field content, but rather to showcase explicit examples embodying these characteristics. Therefore, for the sake of simplicity, we will make several assumptions along our analysis. 

We start by assuming that the extra fermions appear vector-like with respect to the SM gauge group but chiral with respect to the extra $U(1)$. These fermions get Dirac masses through the vacuum expectation value (vev) of a single complex scalar field $S$.  

More precisely, the extra fermions come in representations of $SU(3)\times SU(2)$ as follows:
\begin{itemize}
    \item A set of $2N_{\mathbf L}$ fermions, labeled as $\psi^{{\mathbf L}_i}$, in the $(\mathbf{1},\mathbf{2})$ representation.
    \item A set of $2N_e$ fermions, labeled as $\psi^{e_j}$, in the $(\mathbf{1},\mathbf{1})$ representation.
    \item A set of $2N_d$ fermions, labeled as $\psi^{d_k}$, in the $(\mathbf{3},\mathbf{1})$  and $(\mathbf{\bar 3},\mathbf{1})$ representations.
    \item A set of $2N_{\mathbf Q}$ fermions, labeled as $\psi^{{\mathbf Q}_m}$, in the $(\mathbf{3},\mathbf{2})$ and $(\mathbf{\bar 3},\mathbf{2})$ representations.
\end{itemize}
The complete set of matter fields, along with their associated quantum numbers, is comprehensively displayed in Table~\ref{tab:full_spectrum}.
\begin{figure}[h!]
\begin{center}
\begin{equation*}
\begin{array}{|lllll|cccccc|}
\hline
&&&&&& SU(3)& SU(2)& U(1)_Y&& U(1)_A\\
\textrm{SM sector}&~~&{\mathbf Q}_L^f&&f=1,2,3&& \mathbf{3}& \mathbf{2}& 1/6 &&z_{\mathbf {\mathbf Q}}^f\\
&&u_R^{c,f}&&&& \mathbf{\bar 3}& \mathbf{1}& -2/3 && z_u^f\\
&&d_R^{c,f}&&&& \mathbf{\bar 3}& \mathbf{1}& 1/3 && z_d^f\\
&&{\mathbf L}_L^f&&&& \mathbf{1}& \mathbf{2}& -1/2 && z_{\mathbf L}^f\\
&&e_R^{c,f}&&&& \mathbf{1}& \mathbf{1}& 1 && z_e^f\\
&&\nu_R^{c,f}&&&& \mathbf{1}& \mathbf{1}& 0 && z_\nu^f\\
&&H&&&& \mathbf{1}& \mathbf{2}& 1/2 && z_H\\
\hline
&~~&&&&& & &  && \\
\textrm{Secluded sector}&&\psi_L^{{\mathbf L}_i} &&&& \mathbf{1}& \mathbf{2}& y_{\mathbf L}^i &&q_{{\mathbf L}}^i\\
&&(\psi_R^{{\mathbf L}_i })^c && i=1,\cdots,N_{\mathbf L}&& \mathbf{1}& \mathbf{2}& -y_{\mathbf L}^i &&\widetilde{q_{\mathbf L}}^i\\
&&\psi_L^{e_j} && && \mathbf{1}& \mathbf{1}& y_e^j &&q_e^j\\
&&(\psi_R^{e_j})^c && j=1,\cdots,N_e&& \mathbf{1}& \mathbf{1}& -y_e^j &&\widetilde{q_e}^j\\
&&\psi_L^{d_k} && && \mathbf{3}& \mathbf{1}& y_d^k &&q_d^k\\
&&(\psi_R^{d_k})^c && k=1,\cdots,N_d&& \mathbf{\bar 3}& \mathbf{1}& -y_d^k &&\widetilde{q_d}^k\\
&&{\psi_L^{{\mathbf Q}_m}} && && \mathbf{3}& \mathbf{2}& y_{\mathbf Q}^m &&q_{\mathbf Q}^m\\
&&({\psi_R^{{\mathbf Q}_m}})^c && m=1,\cdots,N_{\mathbf Q} && \mathbf{\bar 3}& \mathbf{2}& -y_{\mathbf Q}^m &&\widetilde{q_{\mathbf Q}}^m\\
&&S&& && \mathbf{1}& \mathbf{1}& 0 && q_S\\
\hline
\end{array}
\end{equation*}
\captionof{table}{The particle content of the model}
\label{tab:full_spectrum}
\end{center}
\end{figure}
Here, the index $f$ spans from 1 to 3 and represents the three SM families.

We assume that the scalar fields 
acquire non-vanishing vacuum expectation values
\begin{equation}
\braket{H}=\frac{1}{\sqrt{2}}
\begin{pmatrix}
    0\\
    v
\end{pmatrix},\quad
\braket{S}=\frac{v_S}{\sqrt{2}}.
\end{equation}
Both of these vev's contribute to the mass $M_A$ of the $U(1)_A$ gauge boson. However, we work in the limit where $v\ll v_S$, where we have
\begin{equation}\label{eq:gauge_boson_mass}
M_A\sim g_A |q_S| v_S.
\end{equation}
Finally, the Yukawa terms $Y_{ij} \bar \psi_L^i \psi_R^i \hat S$, where by $\hat S$ we denote $S$ or $S^*$, give masses to the fermions
\bea
M_{\psi,ij} = Y_{ij} v_S.
\eea
Since we assume that all extra fermions are heavier than energies that can be probed, we consider $Y_{ij}$ to be large enough, typically of order one. Furthermore, for the purposes of our analytical framework, we deem it sufficient to consider the simpler case of $Y_{ij} \propto \delta_{ij}$.

\subsection{Constraints on the charges from Yukawa couplings}
\label{sect:Yukawa_constraints}

The presence of Yukawa couplings among the SM fields establishes the following relationships between the $U(1)_A$ charges of the SM particles:
\bea\label{eq:SM_constraints}
\begin{array}{lllllllllllll}
&~&\bar{{\mathbf Q}}_L^i H d^j_R&~&\to&~&-z_{\mathbf {\mathbf Q}}^i-z_{d}^j+z_H&=&0,\\
&~&\bar{{\mathbf Q}}_L^i\tilde H u_R^j &~&\to&~& -z_{\mathbf {\mathbf Q}}^i-z_{u_R}^j-z_H&=&0,\\
&~&\bar{{\mathbf L}}^iHe_R^j
&~&\to&~& -z_{\mathbf L}^i-z_{e_R}^j+z_H&=&0,
\end{array}
\eea
for $i,j=1,2,3$ the SM families. 

Another constraint arises with the introduction of a Dirac mass term for the neutrinos,
\bea\label{eq:SM_constraint_neutrino}
\begin{array}{lllllllllllll}
&~&\bar{{\mathbf L}}^i\tilde H\nu_R^j&~&\to&~&-z_{\mathbf L}^i-z_{\nu_R}^j-z_H&=&0,
\end{array}
\eea
which, by itself, would necessitate extremely small Yukawa couplings. The popular alternative approach is to consider a see-saw mechanism, achieved by incorporating a Majorana mass term generated through the vev of the field $S$ \cite{PhysRevLett.44.912, Coriano:2015sea, Coriano:2014mpa}:
\bea\label{eq:SM_constraint_neutrino-Majorana}
\begin{array}{lllllllllllll}
\bar \nu_R^{c,i} \nu_R^j \frac{{\hat S}^n}{\Lambda^{n-1}} &~&\to&~& \phantom{-}z_{\nu_R}^i+z_{\nu_R}^j-(\e_\nu^{ij })^n n~q_S&=&0,
\end{array}
\eea
where $\e_\nu^{ij } =\pm 1$ depending if we use $S$ or $S^*$, and $\Lambda$ is a high energy scale which, if to be associated with the vev of an integrated field, this should be neutral  under $U(1)_A$.

The secluded fermions acquire their masses from the vacuum expectation value  of $S$. Initially, we will undertake an analysis similar to that of the SM fermions, introducing the Yukawa coupling at the renormalizable level. Given that the Yukawa interactions can involve either $S$ or $S^*$, we introduce a sign $\e_{\mathbf L}^i, \e_e^j, \e_d^k, \e_{\mathbf Q}^m=\pm 1$. This leads to the following relationships between the $U(1)_A$ charges of the fermions within the secluded sector:
\bea\label{eq:extra_Higgs_constraints}
\begin{array}{lllllllllllll}
&~&\bar \psi_L^{{\mathbf L}_i} \psi_R^{{\mathbf L}_i} \hat S&~&\to&~& -q_{\mathbf L}^i-\widetilde{q_{\mathbf L}}^i+\e_{\mathbf L}^i q_S&=&0,
\\
&~&\bar \psi_L^{e_j} \psi_R^{e_j} \hat S&~&\to&~& -q_e^j-\widetilde{q_e}^j+\e_e^j q_S&=&0,
\\
&~&\bar \psi_L^{d_k} \psi_R^{d_k} \hat S&~&\to&~& -q_d^k-\widetilde{q_d}^k+\e_d^k q_S&=&0,
\\
&~&\bar \psi_L^{{\mathbf Q}_m} \psi_R^{{\mathbf Q}_m} \hat S&~&\to&~& -q_{\mathbf Q}^m-\widetilde{q_{\mathbf Q}}^m+\e_{\mathbf Q}^m q_S&=&0,
\end{array}
\eea
where $\hat S$ denotes either $S$ or $S^*$. 
While we will restrict in the following to the simplest case of renormalizable Yukawa interaction, it is straightforward to generalise the discussion to the case of higher-order non-renormalizable terms as:
\bea\label{eq:extra_Higgs_constraints-non-R}
\begin{array}{lllllllllllll}
&~&\frac{1}{\Lambda^{n-1}}\bar \psi_L^{{\mathbf L}_i} \psi_R^{{\mathbf L}_i} \hat S^n&~&\to&~& -q_{\mathbf L}^i-\widetilde{q_{\mathbf L}}^i+(\e_{\mathbf L}^{i })^n n~  q_S&=&0,
\\
&~&\frac{1}{\Lambda^{m-1}} \bar \psi_L^{e_j} \psi_R^{e_j} \hat S^m&~&\to&~& -q_e^j-\widetilde{q_e}^j+(\e_e^j)^m m~  q_S&=&0,
\\
&~&\frac{1}{\Lambda^{r-1}} \bar \psi_L^{d_k} \psi_R^{d_k} \hat S^r&~&\to&~& -q_d^k-\widetilde{q_d}^k+(\e_d^k)^r r~ q_S&=&0,
\\
&~&\frac{1}{\Lambda^{t-1}} \bar \psi_L^{{\mathbf Q}_m} \psi_R^{{\mathbf Q}_m} \hat S^t&~&\to&~& -q_{\mathbf Q}^m-\widetilde{q_{\mathbf Q}}^m+(\e_{\mathbf Q}^m)^t t~  q_S&=&0,
\end{array}
\eea
where $\Lambda$ is a high energy scale suppressing the non-renormalizable interactions. 


\subsection{Anomaly cancellation conditions}

The conditions to ensure the absence of gauge and mixed gauge-gravitational anomalies within the field content described in Table~\ref{tab:full_spectrum} read:
\bea\label{eq:anomaly_conditions}
\begin{array}{lllllll}
& Tr[Y]_{SM}&= Tr[Y]_{secluded} & = 0,\\
& Tr[YYY]_{SM}&= Tr[YYY]_{secluded} & = 0,\\
& Tr[YT_2T_2]_{SM}&= Tr[YT_2T_2]_{secluded} & = 0,\\
& Tr[YT_3T_3]_{SM}&= Tr[YT_3T_3]_{secluded} & = 0,\\
& Tr[T_3T_3T_3]_{SM}&= Tr[T_3T_3T_3]_{secluded} & = 0,\\
&  &  &  \\
& Tr[q_A]_{SM}&= - Tr[q_A]_{secluded} & \equiv t_{A},\\
& Tr[YYq_A]_{SM}&= - Tr[YYq_A]_{secluded} & \equiv t_{YYA},\\
& Tr[Yq_Aq_A]_{SM}&= - Tr[Yq_Aq_A]_{secluded} & \equiv t_{YAA},\\
& Tr[q_Aq_Aq_A]_{SM}&= - Tr[q_Aq_Aq_A]_{secluded} & \equiv t_{AAA},\\
& Tr[q_AT_2 T_2 ]_{SM}&= -Tr[q_AT_2 T_2 ]_{secluded} & \equiv t_2,\\
& Tr[q_AT_3 T_3 ]_{SM}&= -Tr[q_AT_3 T_3 ]_{secluded} & \equiv t_3.
\end{array}
\eea
The notation $Tr[...]_{SM}$ ($Tr[...]_{secluded}$) denotes the traces over all the fermions in the SM (secluded) sector running in the triangle loop, and $T_2^a$ ($T_3^a$) are the generators of the $SU(2)$ ($SU(3)$) factor of the SM gauge group, normalized as $Tr[T_i^aT_j^b]=\delta^{ab}\delta_{ij}/2$, where $i,j$ run over the non-abelian factors and $a,b$ over their generators. For simplicity, we omit the explicit labeling of these generators.

The five first conditions involving the new fermions are trivially satisfied since these fields are vector-like with respect to the SM gauge group. The aim of this section and the next one is to analyze and provide some solutions to the remaining two sets of conditions.

\subsubsection*{Anomalies from the SM fermions}\label{anomalies from the SM side}

Explicitly, from the field content of Table~\ref{tab:full_spectrum}, the anomalies from the SM sector are given by:
\bea
\begin{array}{lllllll}
& Tr[q_A]_{SM} &=  
\sum_f[6 z_{\mathbf {\mathbf Q}}^f + 3 z^f_u + 3 z^f_d + 2 z_{\mathbf L}^f + z^f_e + z^f_\n]
& = t_{A},\\
& Tr[YYq_A]_{SM} 
&=\sum_f[6 (y^f_{\mathbf Q})^2 z_{\mathbf {\mathbf Q}}^f + 3 (y^f_u)^2 z^f_u + 3 (y^f_d)^2 z^f_d + 2 (y^f_L)^2 z_{\mathbf L}^f + (y^f_e)^2 z^f_e ]& = t_{YYA},\\
& Tr[Yq_Aq_A]_{SM} 
&=\sum_f[6 y^f_{\mathbf Q} (z_{\mathbf {\mathbf Q}}^f)^2 + 3 y^f_u (z^f_u)^2 + 3 y^f_d (z^f_d)^2 + 2 y^f_L (z_{\mathbf L}^f)^2 + y^f_e (z^f_e)^2]& = t_{YAA},\\
& Tr[q_Aq_Aq_A]_{SM} 
&=\sum_f[6 (z_{\mathbf {\mathbf Q}}^f)^3 + 3 (z^f_u)^3 + 3 (z^f_d)^3 + 2 (z_{\mathbf L}^f)^3 + (z^f_e)^3
+ (z^f_\n)^3]
& = t_{AAA},\\
& Tr[q_AT_2T_2]_{SM} &= \sum_f [3 z_{\mathbf {\mathbf Q}}^f + z_{\mathbf L}^f]
 & = t_2,\\
& Tr[q_AT_3T_3]_{SM} &= \sum_f [2 z_{\mathbf {\mathbf Q}}^f+z^f_u+z^f_d] & = t_3,
\end{array}~~~~~~~
\label{eq:SM_anomalies}~~\eea
where the $y^f_{X}$ are the SM fermion $X$ hypercharges. 
%
%
Imposing the Yukawa constraints \eqref{eq:SM_constraints}, the generic formulas in \eqref{eq:SM_anomalies} become:
\bea\label{eq:simplified_SM_anomalies_NOneutrinos}
\begin{array}{llllllllll}
& Tr[q_A]_{SM} &= \sum_f [2z_{\mathbf L}^f+z_e^f+z_{\nu}^f]& = t_{A},\\
& Tr[YYq_A]_{SM} 
&=-{{\frac{1}{2}}}\sum_f[3z_{\mathbf {\mathbf Q}}^f+z_{\mathbf L}^f]
& = t_{YYA},\\
& Tr[Yq_Aq_A]_{SM} 
&=-{{2}} \sum_f [3z_{\mathbf {\mathbf Q}}^f+z_{\mathbf L}^f]z_H& = t_{YAA},\\
& Tr[q_Aq_Aq_A]_{SM} 
&=\sum_f[z_H^3\!+\!3z_H (z_{\mathbf L}^f)^2\!+\!(z_{\mathbf L}^f)^3\! -\! 3z_H^2 z_{\mathbf L}^f\!-\!18 z_H^2z_{\mathbf {\mathbf Q}}^f+(z_{\nu}^f)^3]& = t_{AAA},\\
& Tr[q_AT_2T_2]_{SM} &=\sum_f [3 z_{\mathbf {\mathbf Q}}^f + z_{\mathbf L}^f]
& = t_2,\\
& Tr[q_AT_3T_3]_{SM} &= 0 & = t_3.
\end{array}
\eea
Prior to enforcing the constraints \eqref{eq:SM_constraint_neutrino} stemming from the presence of neutrino Dirac masses, three independent anomalies ($t_A$, $t_{AAA}$, $t_2$) are present and can be chosen as $t_A$, $t_{AAA}$ and $t_2$ while
\bea\label{eq: anomalies after Y without neutrinos}
t_{YYA}=-\frac{1}{2} t_2
~,~~t_{YAA}=-2z_H t_2
~,~~t_3=0.
\eea
By imposing \eqref{eq:SM_constraint_neutrino}, additional relations are introduced, 
\bea\label{eq: anomalies after Y with neutrinos}
t_A=0~,~~t_{AAA}=-6z_H^2 t_2~,
\eea
thereby establishing connections between all SM anomalies. These anomalies can now be expressed as functions of a single anomaly, such as the mixed anomaly $t_2$, and the charge $z_H$.

\subsubsection*{Anomalies from the secluded sector}\label{anomalies from the extra sector side}
The anomaly equations arising from the secluded sector of the field content of Table~\ref{tab:full_spectrum} yield rather lengthy expressions which are presented in Appendix~\ref{app:full_expressions}, in their unconstrained form in Eqs.~\eqref{eq:secluded_anomalies} and after imposing the Yukawa constraints \eqref{eq:extra_Higgs_constraints} in Eqs.~\eqref{eq:simplified_secluded_anomalies}.

To facilitate our analysis and seek specific charge assignments that result in an anomaly-free spectrum for the field content listed in Table~\ref{tab:full_spectrum}, we introduce additional assumptions to simplify the model. Specifically, we assume charge universality with respect to both $U(1)_Y$ and $U(1)_A$ for all secluded particles of each type: ${\mathbf L}$, $e$, ${\mathbf {\mathbf Q}}$, and $d$. In other words, we assume the following:
\bea\label{simple charge assignements}
\ba{llllllllllll}    
&\forall ~i=1,...,N_{\mathbf L} &~~~~~& \forall ~j=1,...,N_e &~~~~~& \forall ~k=1,...,N_d &~~~~~& \forall ~m=1,...,N_{\mathbf Q} & \\
&	\e_{\mathbf L}^i = \e_{\mathbf L} &&
	\e_e^j = \e_e &&
	\e_d^k = \e_d &&
	\e_{\mathbf Q}^m = \e_{\mathbf Q} \\
&	y_{\mathbf L}^i = y_{\mathbf L} &&
	y_e^j = y_e &&
	y_d^k = y_d &&
	y_{\mathbf Q}^m = y_{\mathbf Q} \\
&    q_{\mathbf L}^i = q_{\mathbf L} &&
	q_e^j = q_e &&
	q_d^k = q_d &&
	q_{\mathbf Q}^m = q_{\mathbf Q} \\
&   \widetilde{q_{\mathbf L}}^i = \widetilde{q_{\mathbf L}} &&
	\widetilde{q_e}^j = \widetilde{q_e} &&
	\widetilde{q_d}^k = \widetilde{q_d} &&
	\widetilde{q_{\mathbf Q}}^m = \widetilde{q_{\mathbf Q}}
\ea .
\eea
Under these assumptions, the anomaly conditions given in Eqs.~\eqref{eq:simplified_secluded_anomalies} can be simplified to:
\bea
\begin{array}{lllllll}
Tr[q_A]_{secluded} &= \Big(2\e_{\mathbf L} N_{\mathbf L}+\e_e N_e+3\e_dN_d + 6\e_{\mathbf Q} N_{\mathbf Q} \Big) q_S & = -t_{A}, \\
Tr[YYq_A]_{secluded} &= \Big(2 \e_{\mathbf L} y_{\mathbf L}^2 N_{\mathbf L} +\e_e y_e^2 N_e
\\
& ~~~~ +3 \e_d y_d^2 N_d 
+6 \e_{\mathbf Q} y_{\mathbf Q}^2 N_{\mathbf Q} \Big) q_S & = -t_{YYA},\\
Tr[Yq_Aq_A]_{secluded} &= -q_S^2 \Big(2y_{\mathbf L} N_{\mathbf L}+y_e N_e+3 y_d N_d+6 y_{\mathbf Q} N_{\mathbf Q} \Big)\\
& ~~~ +2 q_S \Big(2 \e_{\mathbf L} y_{\mathbf L} q_{\mathbf L} N_{\mathbf L}+\e_e y_e q_e N_e\\
& ~~~~~~~~~~~~ +3\e_d y_d q_d N_d +6\e_{\mathbf Q} y_{\mathbf Q} q_{\mathbf Q} N_{\mathbf Q} \Big) & = -t_{YAA},\\
Tr[q_Aq_Aq_A]_{secluded} &= q_S^3\Big(2\e_{\mathbf L} N_{\mathbf L}+\e_e N_e+3\e_d N_d+6\e_{\mathbf Q} N_{\mathbf Q} \Big) & \\
& ~~ -3q_S^2 \Big(2 q_{\mathbf L} N_{\mathbf L}+q_e N_e+3 q_d N_d+6 q_{\mathbf Q} N_{\mathbf Q} \Big)\\
& ~~ +3 q_S \Big(2\e_{\mathbf L} q_{\mathbf L}^2 N_{\mathbf L}+\e_e q_e^2 N_e
\\
& ~~~~~~~~~~~ +3\e_d q_d^2 N_d + 6\e_{\mathbf Q} q_{\mathbf Q}^2 N_{\mathbf Q} \Big) & = -t_{AAA},\\
Tr[q_AT_2T_2]_{secluded} &= (\e_{\mathbf L} N_{\mathbf L}+3\e_{\mathbf Q} N_{\mathbf Q})q_S  & = -t_2, & 
\\
Tr[q_AT_3T_3]_{secluded} &= (\e_d N_d+2\e_{\mathbf Q} N_{\mathbf Q})q_S & = -t_3. 
\end{array}
\eea
%
In the subsequent analysis, we will determine charge assignments that fulfill the aforementioned relations.

\section{Anomaly-free solutions}
\label{sect:anomaly_free_sol}

Our primary aim is not to provide an exhaustive classification of all conceivable solutions, which would be intricate and of limited utility at this stage. Instead, our goal is to demonstrate the existence of viable solutions to this problem and highlight some of their features. In particular, we present a collection of explicit solutions where all additional family numbers $N_{{\mathbf L}/e/{\mathbf Q}/d}$ are non-zero (Subsection~\ref{sec:all N non zero}). In addition, we present two simple solutions characterized by $(N_{\mathbf Q},N_d)$ or $(N_{\mathbf L},N_e)$ being zero in Subsection~\ref{sec:some N=0}.

\subsection{Solution with all \texorpdfstring{$N_i$}'s non-vanishing}
\label{sec:all N non zero}

The equations $t_3=0$ and $t_A=0$ lead to:
\begin{subequations}\label{eq:tA_t3}
\begin{eqnarray}
2\e_{\mathbf Q} N_{\mathbf Q}+\e_d N_d&=&0,\\
2\e_{\mathbf L} N_{\mathbf L}+\e_e N_e&=&0.
\end{eqnarray}
\end{subequations}
Since the $N_i s$ are positive integers, we must have $\e_{\mathbf L} \e_e=-1$, $\e_{\mathbf Q} \e_d=-1$ and thus
\begin{equation}\label{eq:N_relations}
N_e=2N_{\mathbf L}, \quad N_d=2 N_{\mathbf Q}.
\end{equation}
Using these two relations, the equations for $t_{YYA}$, $t_{YAA}$ and $t_{AAA}$ yield the following system:
\begin{subequations}\label{eq:general_system_massive_RH_neutrinos}
\begin{eqnarray}
\label{eq:t_YYA}
&&2\e_{\mathbf L} N_{\mathbf L}\left[y_{\mathbf L}^2-y_e^2\right]+6\e_{\mathbf Q} N_{\mathbf Q}\left[y_{\mathbf Q}^2-y_d^2\right]=\frac{t_2}{2 q_S},\\
&&
-2q_S^2\left[N_{\mathbf L}(\! y_{\mathbf L}\!+\!y_e\!)\!+\!3N_{\mathbf Q}(y_d\!+\!y_{\mathbf Q})\right]\!+\!4q_S\left[\e_{\mathbf L} N_{\mathbf L}(y_{\mathbf L} q_{\mathbf L}\!-\!y_e q_e)\!+\!3\e_{\mathbf Q} N_{\mathbf Q}(y_{\mathbf Q} q_{\mathbf Q}\!-\!y_d q_d)\right]\!=\!2z_H t_2 \nonumber,\\
&&
 \label{eq:t_YAA}
\\
&&
  \label{eq:t_AAA}
-q_S^2 \left[N_{\mathbf L} (q_{\mathbf L}\!+\!q_e)\!+\!3N_{\mathbf Q}(q_d\!+\!q_{\mathbf Q}) \right]+ q_S\left[ \e_{\mathbf L} N_{\mathbf L}(q_{\mathbf L}^2\!-\!q_e^2)\!+\!3\e_{\mathbf Q} N_{\mathbf Q}(q_{\mathbf Q}^2\!-\!q_d^2)\right] = z_H^2t_2.
\end{eqnarray}
\end{subequations}

The $t_2$ anomaly equation links $N_{\mathbf L}$ and $N_{\mathbf Q}$ as follows:
\begin{equation}\label{eq:t2}
\e_{\mathbf L} N_{\mathbf L}+3\e_{\mathbf Q} N_{\mathbf Q}=-\frac{t_2}{q_S}.
\end{equation}

We solve them by requiring the following:
\begin{itemize}
\item All charges are rational numbers.
\item The lepton-like extra fermions $\psi^{\mathbf L}$ have electric charges $0$ or $\pm 1$, and $\psi^e$ electric charge $\pm 1$.
\item The quark-like extra fermions $\psi^{\mathbf Q}$ and $\psi^d$ have electric charges $\pm 1/3$ or $\pm 2/3$. Indeed, this condition ensures that when the color forces confine, the resulting bound states can all carry integer charges.
\item  We will consider $\e_{\mathbf L}=1$, $\e_e=-1$, $\e_d=1$ and $\e_{\mathbf Q}=-1$.
\end{itemize}
We end up with $12$ parameters: $y_{\mathbf L}$, $y_e$, $y_{\mathbf Q}$, $y_d$, $q_{\mathbf L}$, $q_e$, $q_{\mathbf Q}$, $q_d$, $q_S$, $z_H$, $N_{\mathbf L}$ and $N_{\mathbf Q}$ for the three equations \eqref{eq:general_system_massive_RH_neutrinos}.
The above requirements lead to the hypercharges\footnote{Another possibility for the hypercharge of $\psi^d$ would be $y_d=\pm 1/3$. However, this gives $y_e=\pm\sqrt{\frac{N_{\mathbf L}-N_{\mathbf Q}}{2N_{\mathbf L}}}$, which vanishes for $N_{\mathbf L}=N_{\mathbf Q}$. This case is excluded, since it leads to extra fermions $\psi^e$ not charged under the SM gauge group. On the other hand, $y_d=\pm 2/3$ gives $y_e=\pm\sqrt{\frac{N_{\mathbf L}+N_{\mathbf Q}}{2N_{\mathbf L}}}$, which is still rational when $N_{\mathbf L}=N_{\mathbf Q}$ and gives $y_e=\pm 1$.}
\begin{equation}\label{eq:ychargessolutions}
y_{\mathbf L}=\pm \frac{1}{2},\quad y_{\mathbf Q}=\pm\frac{1}{6}, \quad y_d=\pm\frac{2}{3},\quad y_e=\pm 1,
\end{equation}
together with\footnote{Rationality of the $y_e$ charge can also be obtained with other particle numbers assignments, such that $N_{\mathbf L}=8 N_{\mathbf Q}$ or $N_{\mathbf L}=49 N_{\mathbf Q}$, but this leads to fractional electric charge for $\psi^e$.} 
\be\label{eq:N_relation2}
N_{\mathbf L}=N_{\mathbf Q}.
\ee
This charge assignment directly solves the equation  \eqref{eq:t_YYA}. Since $N_{\mathbf L}=N_{\mathbf Q}$, $N_{\mathbf L}$ simplifies in the two remaining equations \eqref{eq:t_YAA}-\eqref{eq:t_AAA}, so that we end with $6$ parameters $q_{\mathbf L}$, $q_e$, $q_{\mathbf Q}$, $q_d$, $q_S$ and $z_H$ for two equations.

To obtain rational solutions for the $U(1)_A$ charges $q$ within the secluded sector, one sufficient condition we have chosen is to eliminate the quadratic terms in Equation \eqref{eq:t_AAA} by imposing $q_{\mathbf L}=\pm q_e$ and $q_{\mathbf Q}=\pm q_d$. These conditions lead to three distinct possibilities:\footnote{It is easy to see from \eqref{eq:t_AAA} that $q_{\mathbf L}=-q_e$ and $q_{\mathbf Q}=-q_d$ cannot be imposed simultaneously since $t_2\neq 0$.}
\begin{subequations}\label{eq:qcharges_solutions}
\be
\!\!\!\!\!\textrm{Case 1:}~~~~~ \left\{\!\begin{array}{lll}
\dis q_{\mathbf L}& \dis = q_e & \dis =\frac{2q_Sz_H+q_S^2\left[y_{\mathbf L}+y_e+3(y_d+y_{\mathbf Q})\right]-2z_H^2(y_{\mathbf Q}-y_d)}{2q_S(y_{\mathbf L}-y_e+y_{\mathbf Q}-y_d)},\\
\dis q_{\mathbf Q}& \dis = q_d & \dis =\frac{2q_Sz_H+q_S^2\left[y_{\mathbf L}+y_e+3(y_d+y_{\mathbf Q})\right]+2z_H^2(y_{\mathbf L}-y_e)}{-6q_S(y_{\mathbf L}-y_e+y_{\mathbf Q}-y_d)}.\esp
\end{array}\right.
\label{eq:qcharges_solutions_1}
\ee
\be
\textrm{Case 2:}~~~~~ \left\{\!\begin{array}{lll}
\dis q_{\mathbf L}& \dis = -q_e & \dis =\frac{2q_Sz_H+q_S^2\left[y_{\mathbf L}+y_e+3(y_d+y_{\mathbf Q})\right]-2z_H^2(y_{\mathbf Q}-y_d)}{2q_S(y_{\mathbf L}+y_e)},\\
\dis q_{\mathbf Q}& \dis = q_d & \dis =-\frac{z_H^2}{3q_S}.\esp
\end{array}\right.
\label{eq:qcharges_solutions_2}
\ee
\be
\textrm{Case 3:}~~~~~ \left\{\!\begin{array}{lll}
\dis q_{\mathbf L}& \dis = q_e & \dis =-\frac{z_H^2}{q_S},\\
\dis q_{\mathbf Q}& \dis = -q_d & \dis =\frac{2q_Sz_H+q_S^2 \left[y_{\mathbf L}+y_e+3(y_d+y_{\mathbf Q})\right]+2z_H^2(y_{\mathbf L}-y_e)}{-6q_S(y_{\mathbf Q}+y_d)}.\esp
\end{array}\right.
\label{eq:qcharges_solutions_3}
\ee
\end{subequations}
In these expressions, $q_S$ and $z_H$ are considered as free parameters, and the charges denoted by $y$ can assume any of the eight possible values listed in Equation \eqref{eq:ychargessolutions}, excluding cases that result in a vanishing denominator. The charge assignments specified in Equation \eqref{eq:ychargessolutions} combined with those in Equation \eqref{eq:qcharges_solutions} offer a variety of rational solutions for the $U(1)_Y$ and $U(1)_A$ charges. These solutions render the model entirely free of anomalies, while the SM and secluded sectors each retain their own (opposite) anomalies.

The different contributions to the anomaly of $U(1)_A$ depend on $z_H$, $q_S$ and $N_{\mathbf Q}$ according to:
\bea\label{eq:t}
t_{YYA}=-q_S N_{\mathbf Q}
~,~~t_{YAA}=-4z_H q_S N_{\mathbf Q}
~,~~t_{AAA}=-12z_H^2 q_S N_{\mathbf Q}~,~~t_2=2q_S N_{\mathbf Q}.
\eea

Upon imposing the Yukawa conditions, we are left with just two independent $U(1)_A$ charges associated with the SM fermions, which we can chose to be $z_{\mathbf Q}^f$ and $z_{\mathbf L}^f$. These charges are related to the anomaly $t_2$ through the equation:
\be\label{eq:relation_t2_SM_charges}
t_2=\sum_{f=1,2,3}(3z_{\mathbf {\mathbf Q}}^f+z_{\mathbf L}^f).
\ee
Assuming charge universality within the three SM families, the above relation, combined with Equation \eqref{eq:t}, results in:
\be
3(3z_{\mathbf Q}+z_{\mathbf L})=2q_SN_{\mathbf Q}.
\ee
The anomaly-free solutions presented in this section are thus characterized by four free parameters:
\bi
\item $q_S$ and $z_H$, whose selection determines the $U(1)_A$ charges of the secluded sector as outlined in Equation \eqref{eq:qcharges_solutions}.
\item A choice of 2 parameters among $N_{\mathbf Q}$, $z_{\mathbf Q}$, and $z_{\mathbf L}$. 
\ei

We present in Table~\ref{tab:general_sol}
\begin{figure}[h!]
\begin{center}
\begin{equation*}
\begin{array}{|lllll|cccccc|}
\hline
&&&&&& SU(3)& SU(2)& U(1)_Y&& U(1)_A\\
&~~&{\mathbf Q}_L^f&&f=1,2,3&& \mathbf{3}& \mathbf{2}& 1/6 &&z_{\mathbf {\mathbf Q}}^f\\
&&u_R^{c,f}&&&& \mathbf{\bar 3}& \mathbf{1}& -2/3 && -z_H-z_{\mathbf {\mathbf Q}}^f\\
&&d_R^{c,f}&&&& \mathbf{\bar 3}& \mathbf{1}& 1/3 && z_H-z_{\mathbf {\mathbf Q}}^f\\
&&{\mathbf L}_L^f&&&& \mathbf{1}& \mathbf{2}& -1/2 && z_{\mathbf L}^f\\
&&e_R^{c,f}&&&& \mathbf{1}& \mathbf{1}& 1 && z_H-z_{\mathbf L}^f\\
&&\nu_R^{c,f}&&&& \mathbf{1}& \mathbf{1}& 0 && -z_H-z_{\mathbf L}^f\\
&&H&&&& \mathbf{1}& \mathbf{2}& 1/2 && z_H\\
\hline
&&&&&&  &  &  &&  \\
&&\psi_L^{{\mathbf L}_i} &&&& \mathbf{1}& \mathbf{2}& -1/2 && 1/2\\
&&(\psi_R^{{\mathbf L}_i })^c && i=1,\cdots,N_{\mathbf L}&& \mathbf{1}& \mathbf{2}& +1/2 && 1/2\\
&&\psi_L^{e_j} && && \mathbf{1}& \mathbf{1}& +1 && 1/2 \\
&&(\psi_R^{e_j})^c && j=1,\cdots,2N_{\mathbf L}&& \mathbf{1}& \mathbf{1}& -1 && -3/2\\
&&\psi_L^{d_k} && && \mathbf{3}& \mathbf{1}& -2/3 && -1/2\\
&&(\psi_R^{d_k})^c && k=1,\cdots,2N_{\mathbf L}&& \mathbf{\bar 3}& \mathbf{1}& +2/3 && 3/2\\
&&{\psi_L^{{\mathbf Q}_m}} && && \mathbf{3}& \mathbf{2}& +1/6 && -1/2\\
&&({\psi_R^{{\mathbf Q}_m}})^c && m=1,\cdots,N_{\mathbf L} && \mathbf{\bar 3}& \mathbf{2}& -1/6 && -1/2\\
&&S&& && \mathbf{1}& \mathbf{1}& 0 && 1\\
\hline
\end{array}
\end{equation*}
\captionof{table}{Anomaly-free solution, with all $N_i s\neq 0$, with $y_{\mathbf L}=-1/2, y_{\mathbf Q}=1/6, y_d=-2/3, y_e=1, q_S=z_H=1$, in the Case 1 \eqref{eq:qcharges_solutions_1} where $q_{\mathbf L}=q_e, q_d=q_{\mathbf Q}$.}
\label{tab:general_sol}
\end{center}
\end{figure}
an explicit anomaly-free spectrum, keeping general $U(1)_A$ charges $z$ of the SM fermions compatible with the Yukawa conditions. Here, we chose $y_{\mathbf L}=-1/2, y_{\mathbf Q}=1/6, y_d=-2/3, y_e=1, q_S=z_H=1$, in the Case 1 where $q_{\mathbf L}=q_e, q_d=q_{\mathbf Q}$. Several other anomaly-free spectra, with explicit $U(1)_A$ charges of the SM fermions, are presented in Table~\ref{tab:general_sol_with_values},
\begin{figure}[h!]
\begin{center}
\begin{equation*}\label{eq:spectrum_sol_3}
\begin{array}{|lllll|ccc|ccc|ccc|ccc|}
\hline
&& && && && \text{Model $a$} & && \text{Model $b$} & &&  \text{Model $c$} &&\\
\hline
&&&&&& SU(3)& SU(2)& U(1)_Y&& U(1)_A& U(1)_Y&& U(1)_A& U(1)_Y&& U(1)_A\\
&~~&{\mathbf Q}_L^f&&f=1,2,3&& \mathbf{3}& \mathbf{2}& 1/6 && 1/3& 1/6 && 2/3& 1/6 && 1/3\\
&&u_R^{c,f}&&&& \mathbf{\bar 3}& \mathbf{1}& -2/3 && -10/3& -2/3 && -8/3& -2/3 && -4/3\\
&&d_R^{c,f}&&&& \mathbf{\bar 3}& \mathbf{1}& 1/3 && 8/3& 1/3 && 4/3& 1/3 && 2/3\\
&&{\mathbf L}_L^f&&&& \mathbf{1}& \mathbf{2}& -1/2 && 1& -1/2 && 2& -1/2 && 1\\
&&e_R^{c,f}&&&& \mathbf{1}& \mathbf{1}& 1 && 2& 1 && 0& 1 && 0\\
&&\nu_R^{c,f}&&&& \mathbf{1}& \mathbf{1}& 0 && -4& 0 && -4& 0 && -2\\
&&H&&&& \mathbf{1}& \mathbf{2}& 1/2 && 3& 1/2 && 2& 1/2 && 1\\
\hline
&& && && &&  & &&  & &&   &&
\\
 &&\psi_L^{{\mathbf L}_i} &&&& \mathbf{1}& \mathbf{2}& -1/2 && -3& -1/2 && -3& -1/2 && -1\\
&&(\psi_R^{{\mathbf L}_i })^c && i=1,\cdots,N_{\mathbf L}&& \mathbf{1}& \mathbf{2}& +1/2 && 6& +1/2 && 5& +1/2 && 2\\
&&\psi_L^{e_j} && && \mathbf{1}& \mathbf{1}& -1 && -3 & +1 && -3& -1 && -1\\
&&(\psi_R^{e_j})^c && j=1,\cdots,2N_{\mathbf L}&& \mathbf{1}& \mathbf{1}& +1 && 0& -1 && 1& +1 && 0\\
&&\psi_L^{d_k} && && \mathbf{3}& \mathbf{1}& -2/3 && 0& 2/3 && 1/3& -2/3 && 0\\
&&(\psi_R^{d_k})^c && k=1,\cdots,2N_{\mathbf L}&& \mathbf{\bar 3}& \mathbf{1}& +2/3 && 3& -2/3 && 5/3& 2/3 && 1\\
&&{\psi_L^{{\mathbf Q}_m}} && && \mathbf{3}& \mathbf{2}& +1/6 && 0& +1/6 && 1/3& +1/6 && 0\\
&&({\psi_R^{{\mathbf Q}_m}})^c && m=1,\cdots,N_{\mathbf L} && \mathbf{\bar 3}& \mathbf{2}& -1/6 && -3& -1/6 && -7/3& -1/6 && -1\\
&&S&& && \mathbf{1}& \mathbf{1}& 0 && 3& 0 && 2& 0 && 1\\
\hline
&& && && &&  & &&  & &&   &&\\
\hline
&&t_{2}&& && && 6& && 12& && 6 &&\\
&&t_{YYA}&& && && -3& && -6& && -3 &&\\
&&t_{YAA}&& && && -36& && -48& && -12 &&\\
&&t_{AAA}&& && && -324& && -288& && -36 && \\
\hline
\end{array}
\end{equation*}
\captionof{table}{Examples of anomaly-free solutions, with all $N_i s\neq 0$. Model $a$: Anomaly-free solution with $q_{\mathbf L}=q_e, q_d=q_{\mathbf Q}$, and $N_{\mathbf L}=1$, $z_L=1$.
Model $b$: Anomaly-free solution with $q_{\mathbf L}=q_e, q_d=q_{\mathbf Q}$, and $N_{\mathbf L}=3$, $z_L=2$. 
Model $c$: Anomaly-free solution with $q_{\mathbf L}=q_e, q_d=q_{\mathbf Q}$, and $N_{\mathbf L}=3$, $z_L=1$.}
\label{tab:general_sol_with_values}
\end{center}
\end{figure}
together with the numerical values of the non-vanishing anomaly coefficients.\footnote{The spectra presented in Table~\ref{tab:general_sol_with_values} can be made compatible with the presence of a Majorana mass term \eqref{eq:SM_constraint_neutrino-Majorana} for the right-handed neutrino. For this purpose, one can take for instance:
\bi
\item Model $a$: $n=2$, $\e_{\nu}=1$, $z_L=-6$, $z_{\mathbf Q}=8/3$, which implies $z_u=-17/3$, $z_d=1/3$, $z_e=9$, $z_{\nu}=3$.
\item Model $b$: $n=1$, $\e_{\nu}=-1$, $z_L=-1$, $z_{\mathbf Q}=5/3$, which implies $z_u=-11/3$, $z_d=1/3$, $z_e=3$, $z_{\nu}=-1$.
\item Model $c$: $n=2$, $\e_{\nu}=+1$, $z_L=-2$, $z_{\mathbf Q}=4/3$, which implies $z_u=-7/3$, $z_d=-1/3$, $z_e=1$, $z_{\nu}=1$.
\ei
}  

\subsection{Solution with some vanishing \texorpdfstring{$N_i$}'s}
\label{sec:some N=0}

Let us briefly discuss the scenario where certain quantum numbers of the extra fermions are absent. As indicated by Equation \eqref{eq:tA_t3}, meaningful possibilities arise when either $(N_d,N_{\mathbf Q})=(0,0)$ or $(N_{\mathbf L},N_e)=(0,0)$. We will discuss each of these cases in sequence.

$\bullet$ $N_d=N_{\mathbf Q}=0$ case:

This trivially solves $t_3=0$, while the equations for $t_2$ and $t_A$ lead to:
\be
N_e=2N_{\mathbf L}=-2\frac{t_2}{q_S}.
\ee
On the other hand, the equations for $t_{YYA}$, $t_{YAA}$ and $t_{AAA}$ are given by:
\begin{subequations}\label{eq:system_massive_RH_neutrinos}
\begin{eqnarray}
&&y_e^2-y_{\mathbf L}^2=1/4,\\
&&2\e_{\mathbf L} q_S(y_{\mathbf L}+y_e)-4(y_{\mathbf L} q_{\mathbf L}-y_e q_e)-2z_H=0,\\
&&\e_{\mathbf L} q_S(q_{\mathbf L}+q_e)+q_e^2-q_{\mathbf L}^2-z_H^2=0.
\end{eqnarray}
\end{subequations}
These anomaly conditions yield the following relationships between the charges:
\begin{subequations}\label{eq:charges_sol_massive_RH_neutrinos}
\begin{eqnarray}
4 z_H y_e&=&2q_e+q_S,\\
4 z_H y_{\mathbf L}&=&\pm \sqrt{(2q_e+q_S)^2-4 z_H^2},\\
2q_{\mathbf L}^l&=&q_S\pm\sqrt{(2q_e+q_S)^2-4 z_H^2}.
\end{eqnarray}
\end{subequations}
Reality of these solutions imposes the relation $|2q_e+q_S|\geq 2|z_H|$. In this case, one sees that the three conditions considered in Section~\ref{sec:all N non zero} cannot be imposed simultaneously. For instance, considering $y_{\mathbf L}=\pm 1/2$ leads to an irrational $y_e=1/\sqrt{2}$. In order to keep rational charges, we therefore have to relax the condition on the electric charges, allowing fractional ones. The simplest possibility is obtained with: 
\bea\label{eq:qcharges_solutions_NQ=0}
y_{\mathbf L}=0~,~~~y_e=\pm 1/2~,~~~q_e=\pm z_H-q_S/2~,~~~ q_{\mathbf L}=q_S/2.
\eea
The different contributions to the anomaly of $U(1)_A$ depend on $z_H$, $q_S$ and $N_{\mathbf L}$ according to:
\bea
t_{YYA}=\frac{q_S}{2}N_{\mathbf L}
~,~~t_{YAA}=2z_H q_S N_{\mathbf L}
~,~~t_{AAA}=6z_H^2 q_S N_{\mathbf L}~,~~t_2=-q_S N_{\mathbf L}.
\eea
The anomaly $t_2$ is still related to the charges $z_{\mathbf{Q}}^f$ and $z_{\mathbf{L}}^f$ of the SM fermions through \eqref{eq:relation_t2_SM_charges}, which, assuming charge universality within the three SM families, gives:
\be
3(3z_{\mathbf Q}+z_{\mathbf L})=-q_S N_{\mathbf L}.
\ee
Therefore, this solution is again characterized by four free parameters: $q_S$ and $z_H$, whose selection determines the $U(1)_A$ charges of the secluded sector as outlined in Equation \eqref{eq:qcharges_solutions_NQ=0}, together with a choice of 2 parameters among $N_{\mathbf L}$, $z_{\mathbf Q}$, and $z_{\mathbf L}$. This is summarized in Table~\ref{tab:sol_Nd=NQ=0}.
\begin{figure}[h!]
\begin{center}
\begin{equation*}\label{eq:spectrum_sol_1}
\begin{array}{|lllll|cccccc|}
\hline
&&&&&& SU(3)& SU(2)& U(1)_Y&& U(1)_A\\
&~~&{\mathbf Q}_L^f&&f=1,2,3&& \mathbf{3}& \mathbf{2}& 1/6 &&z_{\mathbf Q}\\
&&u_R^{c,f}&&&& \mathbf{\bar 3}& \mathbf{1}& -2/3 && -z_H-z_{\mathbf {\mathbf Q}}\\
&&d_R^{c,f}&&&& \mathbf{\bar 3}& \mathbf{1}& 1/3 && z_H-z_{\mathbf {\mathbf Q}}\\
&&{\mathbf L}_L^f&&&& \mathbf{1}& \mathbf{2}& -1/2 && -q_S N_{\mathbf L}/3-3z_{\mathbf Q}\\
&&e_R^{c,f}&&&& \mathbf{1}& \mathbf{1}& 1 && z_H+q_S N_{\mathbf L}/3+3z_{\mathbf Q}\\
&&\nu_R^{c,f}&&&& \mathbf{1}& \mathbf{1}& 0 && -z_H+q_S N_{\mathbf L}/3+3z_{\mathbf Q}\\
&&H&&&& \mathbf{1}& \mathbf{2}& 1/2 && z_H\\
\hline
&&&&&&& & &&
\\
&&\psi_L^{{\mathbf L}_i} &&&& \mathbf{1}& \mathbf{2}& 0 && q_S/2\\
&&(\psi_R^{{\mathbf L}_i })^c && i=1,\cdots,N_{\mathbf L}&& \mathbf{1}& \mathbf{2}& 0 &&q_S/2\\
&&\psi_L^{e_j} && && \mathbf{1}& \mathbf{1}& \pm 1/2 && \pm z_H -q_S/2 \\
&&(\psi_R^{e_j})^c && j=1,\cdots,2N_{\mathbf L}&& \mathbf{1}& \mathbf{1}& \mp 1/2 &&\mp z_H-q_S/2\\
&&S&& && \mathbf{1}& \mathbf{1}& 0 && q_S  \\
\hline
\end{array}
\end{equation*}
\captionof{table}{Anomaly-free solution with $(N_d,N_{\mathbf Q})=(0,0)$, $(N_{\mathbf L},N_e)\neq (0,0)$, written in terms of the four free parameters $q_S$, $z_H$, $z_{\mathbf{Q}}$ and $N_{\mathbf{L}}$.}
\label{tab:sol_Nd=NQ=0}
\end{center}
\end{figure}
An anomaly-free spectrum with explicit $U(1)_A$ charges of the SM and secluded fermions is presented in Table~\ref{tab:particular_sol_Nd=NQ=0},
\begin{figure}[h!]
\begin{center}
\begin{equation*}
\begin{array}{|lllll|cccccc|}
\hline
&&&&&& SU(3)& SU(2)& U(1)_Y&& U(1)_A\\
&~~&{\mathbf Q}_L^f&&f=1,2,3&& \mathbf{3}& \mathbf{2}& 1/6 && 1/3\\
&&u_R^{c,f}&&&& \mathbf{\bar 3}& \mathbf{1}& -2/3 && -4/3\\
&&d_R^{c,f}&&&& \mathbf{\bar 3}& \mathbf{1}& 1/3 && 2/3\\
&&{\mathbf L}_L^f&&&& \mathbf{1}& \mathbf{2}& -1/2 && -2\\
&&e_R^{c,f}&&&& \mathbf{1}& \mathbf{1}& 1 && 3\\
&&\nu_R^{c,f}&&&& \mathbf{1}& \mathbf{1}& 0 && 1\\
&&H&&&& \mathbf{1}& \mathbf{2}& 1/2 && 1\\
\hline
&&&&&&  &  &  &&  \\
&&\psi_L^{{\mathbf L}_i} &&&& \mathbf{1}& \mathbf{2}& 0 && 1/2\\
&&(\psi_R^{{\mathbf L}_i })^c && i=1,\cdots,N_{\mathbf L}&& \mathbf{1}& \mathbf{2}& 0 && 1/2\\
&&\psi_L^{e_j} && && \mathbf{1}& \mathbf{1}& +1/2 && 1/2 \\
&&(\psi_R^{e_j})^c && j=1,\cdots,2N_{\mathbf L}&& \mathbf{1}& \mathbf{1}& -1/2 && -3/2\\
&&S&& && \mathbf{1}& \mathbf{1}& 0 && 1 \\
\hline
&&&& && & & && \\ 
\hline
&&t_{2}&& && &  -3&&&  \\
&&t_{YYA}&& && & 3/2&&&\\
&&t_{YAA}&& && & 6&&&\\
&&t_{AAA}&& && & 18&&& \\
\hline
\end{array}
\end{equation*}
\captionof{table}{Example of an anomaly-free solution with $(N_d,N_{\mathbf Q})=(0,0)$, $(N_{\mathbf L},N_e)\neq (0,0)$, with the choice of the free parameters $q_S=1$, $z_H=1$, $z_{\mathbf{Q}}=1/3$ and $N_{\mathbf{L}}=3$. }
\label{tab:particular_sol_Nd=NQ=0}
\end{center}
\end{figure}
together with the numerical values of the non-vanishing anomaly coefficients.

$\bullet$ $N_{\mathbf L}=N_e=0$ case:

The condition $t_A=0$ is trivially solved while the equations for $t_2$ and $t_3$ give:
\begin{equation}
N_d=2N_{\mathbf Q}=\frac{2}{3}\frac{t_2}{q_S}.
\end{equation}
The equations for $t_{YYA}$, $t_{YAA}$, and $t_{AAA}$ mirror those in \eqref{eq:system_massive_RH_neutrinos} with the following replacements:
\begin{eqnarray}
&& y_e\rightarrow y_d,~~
y_{\mathbf L}\rightarrow y_{\mathbf Q},~~
q_{\mathbf L}\rightarrow q_{\mathbf Q},~~
q_e\rightarrow q_d,~~
\e_{\mathbf L}\rightarrow \e_{\mathbf Q},~~
\e_e\rightarrow \e_d.
\end{eqnarray}
This substitution leads to solutions analogous to those in \eqref{eq:charges_sol_massive_RH_neutrinos}, which are summarized in Table~\ref{tab:sol_NL=Ne=0}.
\begin{figure}[h!]
\begin{center}
\begin{equation*}\label{eq:spectrum_sol_2}
\begin{array}{|lllll|cccccc|}
\hline
&&&&&& SU(3)& SU(2)& U(1)_Y&& U(1)_A\\
&~~&{\mathbf Q}_L^f&&f=1,2,3&& \mathbf{3}& \mathbf{2}& 1/6 &&z_{\mathbf Q}\\
&&u_R^{c,f}&&&& \mathbf{\bar 3}& \mathbf{1}& -2/3 && -z_H-z_{\mathbf Q}\\
&&d_R^{c,f}&&&& \mathbf{\bar 3}& \mathbf{1}& 1/3 && z_H-z_{\mathbf {\mathbf Q}}\\
&&{\mathbf L}_L^f&&&& \mathbf{1}& \mathbf{2}& -1/2 && q_S N_{\mathbf{Q}}-3z_{\mathbf Q}\\
&&e_R^{c,f}&&&& \mathbf{1}& \mathbf{1}& 1 && z_H-q_S N_{\mathbf{Q}}+3z_{\mathbf Q}\\
&&\nu_R^{c,f}&&&& \mathbf{1}& \mathbf{1}& 0 && -z_H-q_S N_{\mathbf{Q}}+3z_{\mathbf Q}\\
&&H&&&& \mathbf{1}& \mathbf{2}& 1/2 && z_H\\
\hline
&&&& && & &  && 
\\
&&\psi_L^{d_k} && && \mathbf{3}& \mathbf{1}& \pm 1/2 && \pm z_H-q_S/2\\
&&(\psi_R^{d_k})^c && k=1,\cdots,2N_{\mathbf Q}&& \mathbf{\bar 3}& \mathbf{1}& \mp 1/2 && \mp z_H -q_S/2\\
&&{\psi_L^{{\mathbf Q}_m}} && && \mathbf{3}& \mathbf{2}& 0 && q_S/2\\
&&({\psi_R^{{\mathbf Q}_m}})^c && m=1,\cdots,N_{\mathbf Q} && \mathbf{\bar 3}& \mathbf{2}& 0 && q_S/2\\
&&S&& && \mathbf{1}& \mathbf{1}& 0 && q_S\\
\hline
\end{array}
\end{equation*}
\captionof{table}{Anomaly-free solution with $(N_d,N_{\mathbf Q})\neq (0,0)$, $(N_{\mathbf L},N_e)=(0,0)$, written in terms of the four free parameters $q_S$, $z_H$, $z_{\mathbf{Q}}$ and $N_{\mathbf{Q}}$.}
\label{tab:sol_NL=Ne=0}
\end{center}
\end{figure}
However, in this case, there are additional fermions that transform under the $SU(3)$ of the SM ($\psi^{\mathbf Q}$ and $\psi^d$), which are not of the type $\psi^{\mathbf L}$ and $\psi^e$.


In both scenarios outlined earlier, the unconfined additional fermions possess an electric charge of 1/2. However, the existence of these states in our Universe is highly constrained by stringent bounds. Meeting these constraints necessitates a mechanism that effectively suppresses their production throughout the Universe's history. However, further discussion on this matter is beyond the scope of this work.

\section{The Effective Theory}\label{sect:effective_model}

The comprehensive model outlined above is anomaly-free. However, this is not the case for the effective theory derived at energies below $M_f$ after integrating out both the secluded fermions and one of the two real degrees of freedom of the complex scalar $S$. In this effective theory, the SM fields coexist with the additional vector boson $Z^\prime$ and an axion $a$, corresponding to the second degree of freedom of $S$.  The axionic field $a$ serves as a Goldstone boson and can be absorbed through a field redefinition by $Z^\prime$, rendering the latter massive with three polarization modes.

Below $M_f$ in the effective theory, the contribution from triangle Feynman one-loop diagrams to gauge anomalies may not vanish, leading to non-zero coefficients $t_{ijk}$ from the SM fermions. Superficially, the model appears anomalous. However, after integrating out the secluded sector fermions, effective terms, in particular \emph{axionic} and \emph{generalized Chern-Simons} terms, are generated. These terms play a crucial role in the anomaly cancellation mechanism, and the associated currents are conserved \cite{Anastasopoulos:2006cz}. This mechanism offers a means to generate three-point couplings between the gauge bosons in effective low-energy models.

The bosonic part of the effective Lagrangian reads
\bea
\cal{L}=\cal{L}_{\rm kin}+\cal{L}_{\rm axion}+\cal{L}_{\rm GCS},
\eea
with $\cal{L}_{\rm kin}$ the effective Lagrangian describing the kinetic terms of the fields:
\begin{eqnarray}
\cal{L}_{\rm kin} &=&-\frac{1}{2}\sum_{i=2,3}{\rm Tr}_i[G_{i,\mu\nu}G_i^{\mu\nu}]
-\frac{1}{4} F_{A,\mu \nu} F_A^{\mu \nu}-\frac{1}{4}F_{Y,\mu\nu} F^{\mu\nu}_Y\nn\\
&&\qquad-\frac{1}{2} (\partial_\mu a + M_A A_\mu)^2-(\mathcal{D}_{\mu}H)^{\dagger} (\mathcal{D}_{\mu}H).
\end{eqnarray}
Here $i=2,3$ stand for the $SU(2)$ and $SU(3)$ non-abelian factors of the gauge group, $G_{i,\mu\nu}$ are the corresponding non-abelian field strengths and ${\rm Tr}_i$ the appropriate trace in the fundamental representation. $F_{Y,\mu\nu}$ and $F_{A,\mu\nu}$ are the  field strengths for the hypercharge $Y_{\mu}$ and anomalous $A_{\mu}$, respectively. The axion $a$ can be shifted away to provide a mass $M_A = g_A |q_S| v_S$ to $A_{\mu}$ through a Stueckelberg mechanism. The covariant derivative of the SM Higgs $H$ is given by:
\begin{equation}
\mathcal{D}_{\mu}H=\Big(\partial_{\mu}-i g_2 T_2^a W_{\mu}^a-i g_Y y_H Y_{\mu}-i g_A z_H A_{\mu}\Big)H.
\end{equation}
Notice that after electroweak symmetry breaking where $H$ gets a vev $v$, this leads to an extra contribution to the mass of the anomalous $U(1)_A$, which can be neglected assuming $v\ll v_S$.

As previously mentioned, the tree-level effective Lagrangian also incorporates gauge non-invariant terms, the variations of which cancel the one-loop anomalies. These terms are referred to as the axionic and generalized Chern-Simons (GCS) terms, and they are expressed as:
\begin{eqnarray}
\label{eq:L_axion}
\cal{L}_{\rm axion}&=&
\frac{a}{24\p^2} \Big( c_{YY} F_Y \wedge F_Y + c_{YA} ~F_Y \wedge F_A+ c_{AA} ~F_A \wedge F_A\nn\\
&&\qquad+\sum_{i=2,3} d_i ~{\rm Tr}_i[G_i \wedge G_i] \Big),\\
\label{eq:L_GCS}
\cal{L}_{\rm GCS}&=&
\frac{1}{ 24\p^2}\Big(e_{YAY}
 Y \wedge A \wedge F_Y+ e_{YAA} Y \wedge A \wedge F_A \Big)\nn\\
&&\quad+
\frac{1}{ 24\p^2}
\sum_{i=2,3}f_{i}^A~ A\wedge
{\rm Tr}_i\Big[B_i\wedge \Big(d B_i+\frac{2}{3}B_i\wedge B_i\Big)\Big]\nn\\
&&\qquad+
\frac{1}{ 24\p^2}
\sum_{i=2,3}f_{i}^Y~ Y\wedge
{\rm Tr}_i\Big[B_i\wedge \Big(d B_i+\frac{2}{3}B_i\wedge B_i\Big)
\Big],
\end{eqnarray}
where $B_i$ denote the non-abelian gauge fields, and we have used the notation
\bea
F_{X_1}\wedge F_{X_2}=\frac{1}{4} \e_{\m\n\r\s} F_{X_1}^{\m\n}F_{X_2}^{\r\s},\quad X_1\wedge X_2\wedge F_{X_3}\equiv \frac{1}{2} \e_{\m\n\r\s}X_1^{\m}X_2^{\n}F_{X_3}^{\r\s}
\label{Eq:FF}.
\eea
The coefficients $c$, $d$, $e$, and $f$ are constants that will be determined by enforcing anomaly cancellation, as elaborated upon in the subsequent discussion. Under infinitesimal $U(1)_A$ and $U(1)_Y$ gauge transformations with parameters $\e$ and $\z$, respectively,
\begin{equation}
\delta A_{\mu}=\partial_{\mu}\epsilon,\quad \delta a=-M_A\epsilon,\quad \delta Y_{\mu}=\partial_{\mu}\zeta,
\end{equation}
and non-abelian gauge transformations with parameters $\xi^i$,
\begin{equation}
\delta B_{\mu}^i=D_{\mu}\xi^i=\partial_{\mu}\xi^i+g_i[B_{\mu}^i,\xi^i],\quad \delta G_{\mu\nu}^i=g_i[G_{\mu\nu}^i,\xi^i],
\end{equation}
$\cal{L}_{\rm axion}$ and $\cal{L}_{\rm GCS}$ transform according to: 
\begin{eqnarray}
\label{eq:transfo_L_axion}
\d\cal{L}_{\rm axion}&=&
{-M_A \frac{\e} { 24\p^2}} \Big( c_{YY} F_Y \wedge F_Y + c_{YA} ~F_Y \wedge F_A+ c_{AA} ~F_A \wedge F_A\nn\\
&&\qquad+\sum_{i=2,3} d_i ~{\rm Tr}_i[G_i \wedge G_i] \Big),\\
\label{eq:transfo_L_GCS}
\int \d\cal{L}_{\rm GCS}&=&\frac{1}{24\pi^2}\int\bigg\{-\zeta\Big(e_{YAY} F_A \wedge F_Y+e_{YAA}F_A \wedge F_A+ \sum_{i=2,3} f_i^Y ~{\rm Tr}_i[G_i \wedge G_i] \Big)\nn\\
&+&\epsilon\Big(e_{YAY} F_Y \wedge F_Y + e_{YAA} F_Y \wedge F_A - \sum_{i=2,3} f_i^A ~{\rm Tr}_i[G_i \wedge G_i] \Big)\nn\\
&+&\sum_{i=2,3} \Big( f_i^Y F_Y\wedge {\rm Tr}_i[\xi^i \tilde G_i]+f_i^A F_A\wedge {\rm Tr}_i[\xi^i \tilde G_i]\Big)\bigg\},
\end{eqnarray}
where $\tilde G_i$ denotes the abelian part of $G_i$. On the other hand, the anomalous triangle diagrams yield a non-invariance of the one-loop effective action, given by \cite{Anastasopoulos:2006cz}:
\begin{eqnarray}\label{dS_1loop}
\int \delta \mathcal{L}_{1-{\rm loop}}&=&-
\frac{1}{ 24\p^2} \int
\bigg\{ \zeta \Big(g_Y g_A^2 t_{YAA}F_A\wedge F_A+2 g_Y^2 g_A t_{AYY}F_A\wedge F_Y\Big)\nn\\
&&+~\epsilon\Big(g_A g_Y^2 t_{AYY} F_Y\wedge F_Y
+2 g_Y g_A^2 t_{YAA}F_A\wedge
F_Y+g_A^3 t_{AAA}F_A\wedge F_A \nn\\
&&+ \sum_{i=2,3} g_i^2 t_i {\rm Tr}_i[G_i\wedge G_i] \Big)+\sum_{i=2,3}2 g_i^2 t_i {\rm Tr}_i[\xi^i \tilde G_i]\wedge F_A \bigg\}.
\end{eqnarray}
Considering the anomalous variation \eqref{eq:transfo_L_axion} and \eqref{eq:transfo_L_GCS} of the tree level effective action, the gauge invariance of the total action
\be
\int\Big(\d \mathcal{L}_{1-{\rm loop}} + \d \mathcal{L}_{\rm axion} + \d \mathcal{L}_{\rm GCS}\Big)=0
\ee
sets the coefficients of the axionic and GCS terms in \eqref{eq:L_axion} and \eqref{eq:L_GCS} according to:
\bea\label{eq:sol_from_gauge_transfo}
\begin{array}{lllllll}
& M_A c_{AA}&= -g_A^3 t_{AAA},\\
& M_A c_{AY} + e_{YAA}&= -2 g_Y g_A^2 t_{YAA},\\
& M_A c_{YY} - e_{YAY}&= -g_Y^2 g_A t_{AYY},\\
& M_A d_i+f_i^A&= -g_i^2 t_i,~\forall i=2,3,\\
&  &  &  \\
& e_{YAY}&= -2g_Ag_Y^2 t_{AYY},\\
& e_{YAA}&= -g_Y g_A^2 t_{YAA},\\
& f_i^Y&= 0,~\forall i=2,3,\\
&  &  &  \\
& f_i^A&= 2g_i^2 t_i,~\forall i=2,3.
\end{array}
\eea
The first set of four equations comes from the $U(1)_A$ variation, the second set of three equations comes from the $U(1)_Y$ variation, and the last equation from the non-abelian variations. This can be simplified into:
\begin{eqnarray}
M_A c_{AA}&=&-g_A^3 t_{AAA},\\
M_A c_{AY}&=&-g_Y g_A^2 t_{YAA},\\
M_A c_{YY}&=&-3 g_Y^2 g_A t_{AYY},\\
M_A d_i&=& -3 g_i^2 t_i,~\forall i=2,3,\\
e_{YAY}&=&-2g_A g_Y^2 t_{AYY},\\
e_{YAA}&=& -g_Y g_A^2 t_{YAA},\\
f_i^Y&=&0,~\forall i=2,3,\\
f_i^A&=&2g_i^2t_i,~\forall i=2,3.
\end{eqnarray}

\section{The ultraviolet cut-off of the effective field theory }
\label{cut-off section}

In the context of this study, we are operating within an effective field theory that explores energies above the mass scale of the anomalous $U(1)_A$ but below the mass scale of some or all of the fermions within the secluded sector. Consequently, the contributions from the detectable fermions to the anomaly do not cancel out, and the $U(1)_A$ appears as anomalous in the effective field theory. We will now explore the regime of validity of this framework, focusing on expectations derived from low-energy considerations to ascertain the location of the ultraviolet cut-off. A previous discussion of predictions for limits on ultraviolet cut-offs by Preskill and Swampland conjectures, as well as by the unitarity of an effective field theory with anomalous $U(1)$, can be found in \cite{Craig:2019zkf}.

\subsection{The heavy fermions scale}

We assume knowledge of the mass $M_A$ and the coupling constant $g_A$ of the $U(1)_A$ gauge boson, denoted as $Z^\prime$, either through theoretical calculations or experimental measurements. The vector boson $Z^\prime$ obtains its mass $M_A$ as the sum of two distinct contributions:
\bi

\item The tree-level mass for the gauge boson $Z^\prime$ arises from the Higgs mechanism when the scalar field $S$ acquires a vacuum expectation value. This contribution to the mass is given by:
\bea
M_A^{(0)}= g_A |q_S| v_S,
\label{tree mass}
\eea
where we chose $v_S$ to be a positive quantity.

\item The predominant radiative contribution to the $U(1)_A$ gauge boson $Z^\prime$ mass originates from the non-cancellation of anomalies in triangular loops. This contribution computed by Preskill in \cite{Preskill:1990fr}
is quoted of order:
\begin{equation}
M_A^{(1)} \simeq \left|\frac{[g_A^3 t_{AAA}^{(light)}+2g_A^2g_Y t_{YAA}^{(light)}+g_Ag_Y^2t_{AYY}^{(light)}+g_Ag_2^2 t_2^{(light)}+g_Ag_3^2t_3^{(light)}]\Lambda_{eff}}{64\pi^3}\right|.
\label{Preskill bound}
\end{equation}
\ei
We denoted $\Lambda_{eff}$ the effective theory ultraviolet cut-off scale. The superscript ${(light)}$ is employed to explicitly indicate that we are considering only contributions to the anomaly coefficients $t_2, t_3, t_{AAA}, \ldots$ from the light fermions which lie within the reach of the effective field theory. As the complete model is anomaly free, we can instead use the heavy secluded fermions with an assumed common mass scale $M_f$, which lies beyond the reach of the effective field theory. These fermions may constitute either part or the entire secluded sector. We will use then $t_{AAA}^{(h)}, t_{YAA}^{(h)}, t_{AYY}^{(h)}, t_2^{(h)}, t_3^{(h)}$, where the superscript ${(h)}$ is employed to explicitly indicate that we are considering only contributions to the anomaly coefficients  that stem from the existence of secluded fermions. 

The effective theory cut-off scale $\Lambda_{eff}$ will be approximately equal to the mass scale of the heavy fermions, \ie,
\bea
\Lambda_{eff} \simeq M_f.
\eea
Above this scale, these additional fermions must be taken into account in all tree-level and virtual processes. For instance, the diverse contributions from triangular diagrams to the anomaly cancel, and therefore the contribution, described by Equation \eqref{Preskill bound}, to the $Z^\prime$ mass drops out too \cite{Preskill:1990fr}.  The question is therefore which of the two equations \eqref{tree mass} or \eqref{Preskill bound} is more appropriate to deduce approximately this value of the ultraviolet cut-off from the low-energy theory data.

In our framework, the heavy secluded fermion mass originates from the Yukawa coupling to the scalar field $S$, thus:
\bea
M_f\simeq Y_{ij} v_S \simeq v_S.
\eea
The expression given by \eqref{Preskill bound} is sometimes used as the expected UV cut-off of the theory \cite{Craig:2019zkf}. We would like to investigate here when this can be the case in our models. In general, it is significantly subdominant compared to \eqref{tree mass}.  To prevent this from occurring, it is necessary for $g_A$ not to be too small, and for the $U(1)_A$ charges to be sufficiently large so that $t_{AAA}^{(h)}$ provides a substantial contribution. More precisely, this requirement demands that $g_Aq_A$ be large. The other $t$ coefficients are kept smaller because they involve Casimirs for the fundamental representations of $SU(2)$ and $SU(3)$.  This is a necessary but not sufficient condition. This is because for $M_A \ll M_f$ to hold, $g_Aq_S$ must remain significantly smaller than one. Then, the  Yukawa couplings at origin of the chiral fermion masses requires that the two fermionic secluded fields, left and right, have charges of similar absolute size but opposite signs. This results in compensating contributions to $t_{AAA}^{(h)}$.

The ratio of the loop-induced mass with respect of the tree-level one is now of order:
\begin{equation}
\frac{{M_A}^{(1)}}{{M_A}^{(0)}} \simeq \frac{g_A^2 \left|t_{AAA}^{(h)}\right|}{64\pi^3 |q_S|}\qquad  \overrightarrow{{\textrm{all extra fermions heavy} }} \qquad \frac{{M_A}^{(1)}}{{M_A}^{(0)}} \simeq \frac{3 g_A^2 z_H^2  N_{\mathbf L}}{16 \pi^3 }.
\label{Preskill bound all heavy}
\end{equation}

Indeed, the dominance of the anomaly loop-induced mass for the $Z^\prime$ requires that $g_A^2 z_H^2 N_{\mathbf L} \sim 10^3$. To achieve a light $Z^\prime$ with a mass $M_A \ll M_f$, it necessitates a coupling $g_A \ll 1$. This, in turn, implies significantly large charges and/or a large number of fields $N_{\mathbf L}$, especially if we assume $q_S=1$.

One potential resolution to this issue is that only some of the fermions are heavy enough to lie outside the range of the effective field theory. In such a scenario, the expression in Equation \eqref{Preskill bound all heavy} needs to be reevaluated. Let's consider, for instance, that only the $N_{{\mathbf L}}$ fermions $\psi_L^{{\mathbf L}i}$ with large charges $q_{{\mathbf L}}\gg q_S$ are heavy and inaccessible. In this case:
\begin{equation}
\frac{{M_A}^{(1)}}{{M_A}^{(0)}} \simeq \frac{g_A^2 \left|t_{AAA}^{(h)}\right|}{64\pi^3 |q_S|}\qquad  \overrightarrow{N_{{\mathbf L}} \psi_L^{{\mathbf L}} {\textrm{ heavy} }} \qquad \frac{{M_A}^{(1)}}{{M_A}^{(0)}} \simeq \frac{3 g_A^2 q_{{\mathbf L}}^2  N_{\mathbf L}}{32 \pi^3 }.
\label{Preskill bound L heavy}
\end{equation}
Now, it is $g_A^2 q_{{\mathbf L}}^2 N_{\mathbf L}$ that needs to be of order $\sim 10^3$. In this scenario, $N_{\mathbf L}$ and $q_{{\mathbf L}}$ are allowed to be large and the latter could be arranged by choosing a suitably large value for $z_H$ and not too small $g_A$.

Let us focus on the parameter space where the tree-level mass dominates in our models. In this scenario, the ultraviolet cut-off of the theory can be approximated as:
\bea
\Lambda_{eff} \simeq \frac{M_A}{g_A |q_S|}.
\eea
In the case where $g_A$ is hierarchically the smallest coupling in the theory, the magnetic Swampland Conjecture \cite{Arkani-Hamed:2006emk} can be used to put a bound as:
\bea
\Lambda_{eff} \simlt \Lambda_{QG} \simeq g_A M_P \Rightarrow {M_A} \simlt { g_A^2 } |q_S| M_p.
\eea
This provides an upper bound in our scenario on the mass $M_A$ in terms of $q_S$ and $g_A$ so that the model doesn't fall in the Swampland.


\subsection{Running of the gauge couplings}

For the discussed models to maintain validity, it is crucial that all couplings remain perturbative across the entire energy range of the corresponding effective field theory. This requirement entails the following conditions:

\begin{itemize}
    \item In the low-energy effective field theory, the gauge couplings should retain their perturbative nature up to the scale $M_f$ of the heavy fermions.

    \item In the context of the complete anomaly-free model, the couplings must remain perturbative, extending at least to energy scales above $M_f$. One practical approach would be, for example, to ensure their validity for an order of magnitude higher energies beyond this scale.

    \item Alternatively, adopting a `desert-type scenario' perspective, with the potential opportunity for achieving coupling unification, we might seek perturbativity up to very high energies as the Planck scale.

\end{itemize}

Since our goal is a rough estimate of the energy scale at which gauge couplings reach values greater than one, a discussion centered on the one-loop running approximation of their evolution is sufficient. The $\beta$-function associated with the running coupling constant $g_i$ of the factor $i$ of the gauge group is given at one-loop by
\begin{equation}\label{eq:one_loop_beta}
\beta(g_i)\equiv\mu\frac{dg_i}{d\mu}=b_i g_i^3,
\end{equation}
where the constant $b_i$ is:
\begin{equation}
b_i=-\frac{1}{(4\pi)^2}\left[\frac{11}{3}C_2(G_i)-\frac{4}{3}T_F^i-\frac{1}{3}T_{\phi_c}^i\right].
\end{equation}
In this expression, $C_2(G_i)$ is the quadratic Casimir for the gauge group factor $G_i$, while $T_F^i$ and $T_{\phi_c}^i$ are the fermion and complex scalar Dynkin indices. For the cases we are interested in, the Casimir read $C_2(SU(N))=N$ and $C_2(U(1))=0$, while the Dynkin indices are given by $T_F=T_{\phi_c}=1/2$ for a fundamental representation of $SU(N)$, and $T_F=1/2\sum_k (q_{kL}^2+q_{kR}^2)$, $T_{\phi_c}=1/2\sum_k q_{k}^2$, for chiral fermions and complex scalar charged under a $U(1)$. In terms of $\alpha_i\equiv g_i^2/(4\pi)$, equation \eqref{eq:one_loop_beta} can be integrated into
\begin{equation}\label{running of alpha}
\frac{1}{\alpha_i(\mu^2)}=\frac{1}{\alpha_i(\mu_0^2)}-4\pi b_i\ln\frac{\mu^2}{\mu_0^2}~~~~\to ~~~~
\alpha_i(\mu^2)=\frac{\alpha_i(\mu_0^2)}{1-4\pi b_i~\alpha_i(\mu_0^2)~\ln\frac{\mu^2}{\mu_0^2}}.
\end{equation}
Here, $\mu_0$ represents an arbitrary reference scale typically chosen as a known point where the coupling value is established.

In the low-energy effective field theory, the condition that gauge couplings should retain their perturbative nature up to the scale $M_f$ amounts to constrain the value of the $U(1)_A$ coupling $g_A$ at $M_A$. The SM couplings are not affected at these scales. The running of $g_A$ from a measured value at the $Z^\prime$ mass $M_A$ is given by:
\begin{equation}
M_f= M_A \times \exp{\left[\frac{1}{8\pi b_A} \left(\frac{1}{\alpha_A(M_A^2)}- \frac{1}{\alpha_A(M_f^2)}\right)\right]},
\end{equation}
where the beta-function coefficient 
is expected to be of order 1 and is given in \eqref{eq: b's for all gauge groups and ranges}. Clearly, achieving a significant separation between scales $M_A$ and $M_f$ is facilitated by ensuring that $\alpha_A(M_f^2)$ is small. However, the challenge lies in the fact that a small gauge coupling makes it more difficult to produce and detect in future colliders.

We now consider the context of the complete model which has no gauge anomaly. The presence of additional fields alters the running of the couplings at energies beyond their respective masses. Assuming, once again, that all extra fermions and the scalar $S$ acquire mass at the common scale $M_f$, the evolution of the couplings below this scale will exclusively receive contributions from the SM sector. In contrast, both sectors will contribute to the running of the couplings above $M_f$. The gauge couplings at an energy scale $\mu$ therefore read:
\bea
\frac{1}{\alpha_i(\mu^2)}=\frac{1}{\alpha_i(M_A^2)}-
4\pi\left\{
\ba{lllll} b_i\ln\frac{\mu^2}{M_A^2}&~~~~~& M_A<\m<M_f\\
b_i\ln\frac{M_f^2}{M_A^2}+\tilde b_i 
\ln\frac{\mu^2}{M_f^2}&& M_f
\leq \m
\ea\right. .
\eea
The different values for the beta-coefficients $b_i$ and $\tilde b_i$ for the spectrum of Table~\ref{tab:full_spectrum} are given by:
\bea
\ba{|c|ccc|ccc|}
\hline
&&~~~& (4\pi)^2b_i&&~~~& (4\pi)^2\tilde b_i \\
\hline
SU(3) &&& -7 &&& 
-7+\frac{2}{3}(N_d+2N_{\mathbf Q})
\\
SU(2) &&& -19/6 &&& 
-\frac{19}{6}+\frac{2}{3}(N_{\mathbf L}+3N_{\mathbf Q})
\\
U(1)_Y &&& 41/6 &&& 
\frac{41}{6}+\frac{4}{3}\left(2N_{\mathbf L} y_{\mathbf L}^2+N_e y_e^2+3N_d y_d^2+6N_{\mathbf Q} y_{\mathbf Q}^2\right)
\\
U(1)_A &&& \frac{4}{3}T_F^{SM}+\frac{2}{3}z_H^2
 &&& 
\frac{4}{3}(T_F^{\rm SM}+ T_F^{\rm extra}) +\frac{2}{3} z_H^2 + \frac{1}{3}q_S^2
\\
\hline
\ea~~~~
\label{eq: b's for all gauge groups and ranges}\eea
where
\begin{eqnarray}
T_F^{\rm SM}&=&\frac{1}{2}\sum_f\left(6(z_{\mathbf Q}^f)^2+3(z_u^f)^2+3(z_d^f)^2+2(z_{\mathbf L}^f)^2+(z_e^f)^2+(z_{\nu}^f)^2\right),\\
T_F^{\rm extra}&=&N_{\mathbf L}\left[q_{\mathbf L}^2+\widetilde {q_{\mathbf L}}^2\right]+\frac{N_e}{2}\left[q_e^2+\widetilde{q_e}^2\right]+\frac{3}{2}N_d\left[q_d^2+\widetilde{q_d}^2\right]+3N_{\mathbf Q}\left[q_{\mathbf Q}^2+\widetilde{q_{\mathbf Q}}^2\right].
\end{eqnarray}
Using the Yukawa conditions \eqref{eq:SM_constraints} and \eqref{eq:SM_constraint_neutrino}, $T_F^{\rm SM}$ can be simplified into
\begin{equation}\label{eq:TFSM}
T_F^{\rm SM}=2\sum_f\left(3 (z_{\mathbf {\mathbf Q}}^f)^2+(z_{\mathbf L}^f)^2+2z_H^2\right),
\end{equation}
while the Yukawa conditions \eqref{eq:extra_Higgs_constraints} yield:
\begin{eqnarray}\label{eq:TFextra}
T_F^{\rm extra}&=&N_{\mathbf L}\left[2q_{\mathbf L}^2+q_S^2-2\e_{\mathbf L}q_{\mathbf L}q_S\right]+\frac{N_e}{2}\left[2q_e^2+q_S^2-2\e_e q_e q_S\right]\nn\\
&+&\frac{3}{2}N_d\left[2q_d^2+q_S^2-2\e_d q_d q_S\right]+3 N_{\mathbf Q}\left[2q_{\mathbf Q}^2+q_S^2-2\e_{\mathbf Q}q_{\mathbf Q}q_S\right].
\end{eqnarray}
For $\tilde b_i<0$, the theory is asymptotically free and perturbation theory remains trivially valid at high energies. For $\tilde b_i>0$, $\alpha_i(\mu^2)$ increases with $\mu$ and perturbation theory (corresponding to $\alpha_i\simlt 1$) remains valid up to the scale
\begin{equation}\label{eq:scale_breaking_pert}
\Lambda < {\rm min}_i \left( M_f\times \exp\left[\frac{1}{8\pi \widetilde{b_i} \alpha_i(M_f)}\right]\right),
\end{equation}
where ${\rm min}_i$ refers to the minimal value among those obtained by scanning over the gauge group factors $i$.

From now on let us consider the reference scale $\mu_0$ to be at the electroweak scale, $\mu_0=m_Z=91.19~{\rm GeV}$. From the values of the couplings at $m_Z$ \cite{ParticleDataGroup:2022pth},
\begin{equation}\alpha_3(m_Z)=0.1184, \quad \alpha_2(m_Z)=0.0338, \quad  \alpha_Y(m_Z)=0.01014,
\end{equation}
one can use \eqref{running of alpha} to run the couplings up to the scale $M_f$ where the secluded fermions appear. Considering as an example
\bea\label{eq:mu_extra 40TeV}
M_f\sim 40~{\rm TeV},
\eea
we get:
\begin{equation}
\alpha_3(40~{\rm TeV})=0.066, \quad \alpha_2(40~{\rm TeV})=0.0306, \quad  \alpha_Y(40~{\rm TeV})=0.0109.
\end{equation}
With these numerical values, we now consider the running of the SM gauge couplings in the different anomaly-free solutions presented in Section~\ref{sect:anomaly_free_sol}.

\bi
\item Solutions of Section~\ref{sec:all N non zero}, with all $N_i s$ non-vanishing:
\ei
The relations \eqref{eq:N_relations}, \eqref{eq:ychargessolutions} and \eqref{eq:N_relation2} yield
\begin{eqnarray}
(4\pi)^2\tilde b_3&=&-7+\frac{8}{3}N_{\mathbf L},\\
(4\pi)^2\tilde b_2&=&-\frac{19}{6}+\frac{8}{3}N_{\mathbf L},\\
(4\pi)^2\tilde b_Y&=&\frac{41}{6}+\frac{64}{9}N_{\mathbf L}.
\end{eqnarray}
The conditions of asymptotic freedom for $\alpha_3$ and $\alpha_2$ and perturbativity for $\alpha_3$, $\alpha_2$ and $\alpha_Y$ up to $\sim 1000~{\rm TeV}$ or up to (at least) the Planck scale therefore set bounds on the free parameter $N_{\mathbf L}$, as summarised in Table~\ref{tab:bounds_sol_N_nonvanishing}.
\begin{figure}[h!]
\begin{center}
\begin{equation*}\label{eq:1}
\begin{array}{|c|cccc|cc|cc|}
\hline
&&&& i=SU(3)&& i=SU(2) && i=U(1)_Y \\
\hline
\textrm{Asymptotic freedom $\tilde b_i<0$}&&&& N_{\mathbf L} \leq 2 && N_{\mathbf L}=1 &&  - \\
\hline
\textrm{Perturbativity $\alpha_i\leq 1$ up to $1000~{\rm TeV}$}&&&& N_{\mathbf L}\leq 13  && N_{\mathbf L}\leq 20 && N_{\mathbf L} \leq 24 \\
\hline
\textrm{Perturbativity $\alpha_i\leq 1$ up to (at least) $M_{\rm Pl}$}&&&& N_{\mathbf L}\leq 3 && N_{\mathbf L}\leq 3 && N_{\mathbf L}=1 \\
\hline
\end{array}
\end{equation*}
\captionof{table}{Constraints on the number of extra fermions $N_{\mathbf L}$ from the running of the coupling constants, for the anomaly-free solutions of Section~\ref{sec:all N non zero}, with all $N_i s$ non-vanishing.}
\label{tab:bounds_sol_N_nonvanishing}
\end{center}
\end{figure}
%

Considering as an example the particular charge assignment presented in the Model $a$ of Table~\ref{tab:general_sol_with_values} with $N_{\mathbf L}=1$, one can use the expressions of $T_F^{\rm SM}$ and $T_F^{\rm extra}$ given in \eqref{eq:TFSM} and \eqref{eq:TFextra} to compute the running of the $U(1)_A$ coupling. One finds that the $U(1)_A$ coupling remains perturbative up to $\sim 1000~{\rm TeV}$ for any $g_A(40~{\rm TeV})\simlt 0.28$, and up to (at least) the Planck scale for any $g_A(40~{\rm TeV})\simlt 0.09$.

\bi
\item Solution of Section~\ref{sec:some N=0}, with $(N_d, N_{\mathbf Q})=(0,0)$ and $N_e=2N_{\mathbf L}\neq 0$:
\ei
Since no extra fermions are charged under the $SU(3)$, the $\alpha_3$ coupling remains unaffected by the secluded sector: it has the usual asymptotically-free behavior and thus remains perturbative at high energies. The bounds on $N_{\mathbf L}$ set by the conditions of asymptotic freedom for $\alpha_2$ and perturbativity for $\alpha_2$ and $\alpha_Y$ up to $\sim 1000~{\rm TeV}$ or up to (at least) the Planck scale are summarised in Table~\ref{tab:bounds_sol_Nd=NQ=0}.
\begin{figure}[h!]
\begin{center}
\begin{equation*}\label{eq:2}
\begin{array}{|c|cccc|cc|cc|}
\hline
&&&& i=SU(3)&& i=SU(2) && i=U(1)_Y\\
\hline
\textrm{Asymptotic freedom $\tilde b_i<0$}&&&& \forall N_{\mathbf L} && N_{\mathbf L}\leq 4 && - \\
\hline
\textrm{Perturbativity $\alpha_i\leq 1$ up to $1000~{\rm TeV}$}&&&& \forall N_{\mathbf L} && N_{\mathbf L}\leq 100 && N_{\mathbf L}\leq 258 \\
\hline
\textrm{Perturbativity $\alpha_i\leq 1$ up to (at least) $M_{\rm Pl}$}&&&& \forall N_{\mathbf L} && N_{\mathbf L}\leq 14 && N_{\mathbf L}\leq 17 \\
\hline
\end{array}
\end{equation*}
\captionof{table}{Constraints on the number of extra fermions $N_{\mathbf L}$ from the running of the coupling constants, for the anomaly-free solution of Section~\ref{sec:some N=0}, with $(N_d, N_{\mathbf Q})=(0,0)$ and $N_e=2N_{\mathbf L}\neq 0$.}
\label{tab:bounds_sol_Nd=NQ=0}
\end{center}
\end{figure}
The strongest upper bound on $N_{\mathbf L}$ to remain in the perturbative regime is therefore 
\be
N_{\mathbf L}\leq 14,
\ee 
under which perturbativity for the three SM couplings is preserved up to the Planck scale. Any higher value for $N_{\mathbf L}$ will then lower the scale where perturbation theory breaks down.

Considering the particular charge assignment presented in Table~\ref{tab:particular_sol_Nd=NQ=0} with $N_{\mathbf L}=3$, one can again use the expressions of $T_F^{\rm SM}$ and $T_F^{\rm extra}$ to find that the $U(1)_A$ coupling remains perturbative up to $\sim 1000~{\rm TeV}$ for any $g_A(40~{\rm TeV})\simlt 0.6$, and up to (at least) the Planck scale for any $g_A(40~{\rm TeV})\simlt 0.2$.

\bi
\item Solution of Section~\ref{sec:some N=0}, with $(N_{\mathbf L}, N_e)=(0,0)$ and $N_d=2N_{\mathbf Q}\neq 0$:
\ei
The bounds on $N_{\mathbf Q}$ set by the conditions of asymptotic freedom for $\alpha_3$ and $\alpha_2$ and perturbativity for $\alpha_3$, $\alpha_2$ and $\alpha_Y$ up to $\sim 1000~{\rm TeV}$ or up to (at least) the Planck scale are summarised in Table~\ref{tab:bounds_sol_NL=Ne=0}.
\begin{figure}[h!]
\begin{center}
\begin{equation*}\label{eq:3}
\begin{array}{|c|cccc|cc|cc|}
\hline
&&&& i=SU(3)&& i=SU(2) && i=U(1)_Y \\
\hline
\textrm{Asymptotic freedom $\tilde b_i<0$}&&&& N_{\mathbf Q} \leq 2 && N_{\mathbf Q}=1 &&  - \\
\hline
\textrm{Perturbativity $\alpha_i\leq 1$ up to $1000~{\rm TeV}$}&&&& N_{\mathbf Q}\leq 13  && N_{\mathbf Q}\leq 33 && N_{\mathbf Q}\leq 86 \\
\hline
\textrm{Perturbativity $\alpha_i\leq 1$ up to (at least) $M_{\rm Pl}$}&&&& N_{\mathbf Q}\leq 3 && N_{\mathbf Q}\leq 4 && N_{\mathbf Q}\leq 5 \\
\hline
\end{array}
\end{equation*}
\captionof{table}{Constraints on the number of extra fermions $N_{\mathbf Q}$ from the running of the coupling constants, for the anomaly-free solution of Section~\ref{sec:some N=0}, with $(N_{\mathbf L}, N_e)=(0,0)$ and $N_d=2N_{\mathbf Q}\neq 0$.}
\label{tab:bounds_sol_NL=Ne=0}
\end{center}
\end{figure}
The strongest upper bound on $N_{\mathbf Q}$ to remain in the perturbative regime is therefore
\be
N_{\mathbf Q}\leq 3,
\ee
under which perturbativity for the three SM couplings is preserved up to the Planck scale. Any higher value for $N_{\mathbf Q}$ will then lower the scale where perturbation theory breaks down.

We now shift our focus to the Yukawa couplings in the context of the complete anomaly-free model. More precisely, we need to address the issue of the strength of these couplings. In the scenario under consideration, an energy gap has been postulated to exist between the mass scales of the gauge boson $M_A$ and the extra fermions $M_f$, achieved by assuming a hierarchy between the gauge couplings $g_A$ and the Yukawa couplings $Y_{ij}$. For the gauge coupling not to be excessively small -- allowing the direct production of a substantial number of $Z^\prime$ particles in future collider experiments -- the Yukawa coupling must be of order one. However, it is desirable to prevent this coupling from rapidly becoming too large at the scale $M_f$, allowing for an energy range where the theory can be well approximated by the full model without anomalies.

The Renormalization Group Equation (RGE) for the Yukawa couplings $Y_{ij}$ receives positive contributions from $Y_{ij}$, which would lead to their rapid divergence above $M_f$. Since the additional fermions are charged, there are negative contributions proportional to powers of the gauge couplings in the RGE's. Although the coupling $g_A$ is too small to induce significant effects, the contributions from SM gauge couplings are substantial and ensure the presence of non-negligible negative contributions. These negative contributions can prevent Yukawa couplings from becoming non-perturbative immediately above the scale $M_f$. This highlights one of the advantages of our chosen representations for the extra fermions to construct explicit examples of perturbative models.

To conclude this section, let us also mention that perturbative unitarity of the effective theory will also put contraints on the parameters of the model. This comes from the fact that one of the two real degrees of freedom of the complex scalar $S$ has the same mass as the heavy secluded fermions, and therefore decouples from the low energy effective theory. An analysis of perturbative unitarity in the effective theory can be carried out along the lines of \cite{Cui:2017juz,Kahlhoefer:2015bea}. Since the main goal of this paper is to provide a detailed analysis of the anomaly cancellation conditions in the aforementioned setup, we will not perform such unitary analysis, leaving it for future work.

\section{Conclusions}\label{sect: conclusions}

The scenario under consideration is not novel but is well-established and extensively examined in the literature, with numerous details explored. It involves the possibility that, at accessible energies, a subset of chiral fermions is absent. The resulting theory appears anomalous. Starting from an anomaly-free theory, the missing fermions, being too massive to be produced, are integrated out to formulate the low-energy effective theory. A historical example of such a scenario is the Standard Model in the limit of an infinite top quark mass, subject to studies before the experimental discovery of the heavy quark. An important result is that the integration of massive chiral fermions leads to effective interaction terms involving three gauge bosons. These terms ensure the conservation of the current associated with the seemingly anomalous symmetry. This was initially studied in \cite{DHoker:1984izu,DHoker:1984mif}, and later investigated in general detail in \cite{Anastasopoulos:2006cz}. The novelty of our work is to delve deeper into this scenario and propose a realization through explicit models.

Our analysis goes beyond the existing literature on this subject from several aspects. Previous analysis of the anomaly equations in abelian extensions of the Standard Model consider either the anomaly cancellations occurring independently between the SM and secluded sectors (see \eg \cite{Costa:2020krs,Costa:2020dph,Costa:2019zzy}), or an anomaly cancellation between the SM and secluded sectors but without considering a Higgs mechanism giving mass to the secluded fermions \cite{Batra:2005rh}. In addition, the extra fermions in these analysis do not reside in the same representations as the SM quarks and leptons. On the other hand, our analysis considers a set of extra fermions in the $(\mathbf{1},\mathbf{1})$, $(\mathbf{1},\mathbf{2})$, $(\mathbf{3},\mathbf{1})$, $(\mathbf{\bar 3},\mathbf{1})$, $(\mathbf{3},\mathbf{2})$ and $(\mathbf{\bar 3},\mathbf{2})$ representations of $SU(3)\times SU(2)$. Assuming that these extra fermions get their masses from an extra scalar field, we impose the constraints coming from gauge-invariance of the Yukawa terms and then consider the anomaly cancellation to occur between the SM and secluded sectors, and not independently between the two sectors.

In our construction, the anomaly of an abelian $U(1)_A$ gauge group extending the Standard Model is canceled by a set of chiral fermions too massive to be detected. The hierarchy between the masses of these heavy fermions and the $Z^\prime$ boson is explained by the ratio between order one Yukawa and small gauge couplings. One could contemplate a simpler construction where the new heavy fermions have charges only under $U(1)_A$, with otherwise the analysis proceeding along the same lines. The contribution from the $U(1)_A$ gauge coupling being too small, the beta function of the Yukawa coupling is almost entirely controlled by these Yukawa couplings themselves. The beta function is positive and large enough  to make the theory become non-perturbative just above $M_f$. Therefore, such a ``minimalist'' model would lack an energy range for a valid description of its anomaly-free phase. This contrasts with our models. Here, as the new heavy fermions are charged under the Standard Model, their Yukawa coupling beta functions receive sizable negative contributions proportional to (the squares of) the SM gauge couplings. 

In this study, we specifically considered a $Z^\prime$ boson with a multi-TeV mass and utilized it in the assessment of Renormalization Group Equation effects. Alternatively, we could explore the scenario where the $Z^\prime$ boson is light. 
Due to the constraints imposed on new charged or colored particles, the additional fermions would be expected to have masses $M_f$ in the multi-TeV range. Conversely, given the absence of positive search results for extra gauge bosons, we would require 
$Z^\prime$ to be exceptionally weakly coupled, hence very light. The lower limit on its coupling strength might be determined by the Swampland conjectures \cite{Arkani-Hamed:2006emk}, where $M_f$ 
may be identified with the compactification or string scale \cite{Benakli:2020vng,Anchordoqui:2020tlp}.

One generalization of our construction is its supersymmetric extension. This is straightforward. Two additions are necessary. Firstly, we must add a complex scalar partner for each of the extra fermions and a fermionic partner in the adjoint representation, a gaugino, for the gauge boson. These states do not alter the anomaly conditions but modify the RGEs. Additionally, for the scalar field $S$, we must include a chiral fermion. To avoid the emergence of anomalies and to write the Yukawa couplings where we used $S^*$, one can introduce a superfield $S^\prime$ with an opposite charge to that of $S$. Note that in the supersymmetric case, the scalar potential of $S$ is more constrained than in the non-supersymmetric case. In particular, in the absence of couplings with new fields (from a hidden sector breaking supersymmetry, for example), the quartic potential term is now fixed and given by the D-terms. Also, with the choice of extra fermions making new full families Standard Model representations, it is possible to keep unification of the gauge couplings. We leave the investigation of these extensions for future work.

\section*{Acknowledgements}

We would like to thank M. Goodsell for collaboration at the very early stage of this work. 
We also thank X. Chu, S. Cl\'ery, A. Dedes, E. Dudas, H. Eberl, E. Kiritsis, Y. Mambrini and J. Pradler for useful discussions.
F.R. is supported by the Cyprus Research and Innovation Foundation grant EXCELLENCE/0421/0362.
P.A. was supported by FWF Austrian Science Fund via the SAP P 36423-N. I.A. is supported by the Second Century Fund (C2F), Chulalongkorn University, and thanks the IHEP of Austrian Academy of Sciences for hospitality and partial financial support.

\newpage



\appendix
\numberwithin{equation}{section}


\section{Anomaly equations from the secluded sector}
\label{app:full_expressions}

The anomaly equations from the secluded sector of the spectrum presented in Table~\ref{tab:full_spectrum} explicitly read:
\bea\label{eq:secluded_anomalies}
\begin{array}{lllllll}
& Tr[q_A]_{secluded} &=  
2\sum_{i=1}^{N_{\mathbf L}}[q_{\mathbf L}^i+\widetilde{q_{\mathbf L}}^i] + \sum_{j=1}^{N_e}[ q_e^j+\widetilde{q_e}^j] \\
& &~~+
3\sum_{k=1}^{N_k}[q_d^k+\widetilde{q_d}^k]+6\sum_{m=1}^{N_{\mathbf Q}}[q_{\mathbf Q}^m+\widetilde{q_{\mathbf Q}}^m]
& = -t_{A},\\
& Tr[YYq_A]_{secluded} &=
2 \sum_{i=1}^{N_{\mathbf L}}(y^{l_i})^2 [q_{\mathbf L}^{l_i}+q_R^{l_i}]
+\sum_{j=1}^{N_e}(y^{e_j})^2 [q_{\mathbf L}^{e_j}+q_R^{e_j}]
\\
&&~~
+3 \sum_{d=1}^{N_d}(y^{d_k})^2 [q_{\mathbf L}^{d_k}+q_R^{d_k}]+6 \sum_{m=1}^{N_{\mathbf Q}}(y^{{\mathbf Q}_m})^2 [q_{\mathbf L}^{{\mathbf Q}_m}+q_R^{{\mathbf Q}_m}]& = -t_{YYA},\\
& Tr[Yq_Aq_A]_{secluded} &= 
2\sum_{l_i}y^{l_i}[(q_{\mathbf L}^{l_i})^2-(q_R^{l_i})^2]
+\sum_{e_j}y^{e_j}[(q_{\mathbf L}^{e_j})^2-(q_R^{e_j})^2]\\
& &~~
+3\sum_{d_k}y^{d_k}[(q_{\mathbf L}^{d_k})^2-(q_R^{d_k})^2]+6\sum_{{\mathbf Q}_m}y^{{\mathbf Q}_m}[(q_{\mathbf L}^{{\mathbf Q}_m})^2-(q_R^{{\mathbf Q}_m})^2]
& = -t_{YAA},\\
& Tr[q_Aq_Aq_A]_{secluded} &= 
2\sum_{l_i}[(q_{\mathbf L}^{l_i})^3+(q_R^{l_i})^3]+\sum_{e_j}[(q_{\mathbf L}^{e_j})^3+(q_R^{e_j})^3]\\
& &~~+3\sum_{d_k}[(q_{\mathbf L}^{d_k})^3+(q_R^{d_k})^3]+6\sum_{{\mathbf Q}_m}[(q_{\mathbf L}^{{\mathbf Q}_m})^3+(q_R^{{\mathbf Q}_m})^3]
& = -t_{AAA},\\
& Tr[q_AT_2T_2]_{secluded} &= 
\sum_{i=1}^{N_{\mathbf L}}[q_{\mathbf L}^i+\widetilde{q_{\mathbf L}}^i]+ 
3\sum_{m=1}^{N_{\mathbf Q}}[q_{\mathbf Q}^m+\widetilde{q_{\mathbf Q}}^m]& = -t_2,\\
& Tr[q_AT_3T_3]_{secluded} &=
\sum_{k=1}^{N_d}[q_d^k+\widetilde{q_d}^k] + 2 \sum_{m=1}^{N_{\mathbf Q}}[q_{\mathbf Q}^m+\widetilde{q_{\mathbf Q}}^m]& = -t_3.
\end{array}
~~~~~\eea

Assuming that all extra fermions become massive through Yukawa couplings with a single extra Higgs $S$ imposes the conditions \eqref{eq:extra_Higgs_constraints} between the $U(1)_A$ charges of the secluded particles. All the ``secluded anomalies'' \eqref{eq:secluded_anomalies} then depend on the charge $q_S$ and can be rewritten as:
\bea\label{eq:simplified_secluded_anomalies}
\begin{array}{lllllll}
Tr[q_A]_{secluded} &=  
\Big(2\sum_{l_i}\e_{l_i} + \sum_{e_j}\e_{e_j} + 3\sum_{d_k}\e_{d_k} + 6\sum_{{\mathbf Q}_m}\e_{{\mathbf Q}_m} \Big) q_S 
& = -t_{A},\\
Tr[YYq_A]_{secluded} &=
\Big(2 \sum_{l_i}\e_{l_i} (y^{l_i})^2 +\sum_{e_j}\e_{e_j} (y^{e_j})^2 
\\
& ~~~~ +3 \sum_{d_k}\e_{d_k} (y^{d_k})^2 
+6 \sum_{{\mathbf Q}_m}\e_{{\mathbf Q}_m} (y^{{\mathbf Q}_m})^2 \Big) q_S& = -t_{YYA},\\
Tr[Yq_Aq_A]_{secluded} 
& = -q_S^2 \Big(2\sum_{l_i}y^{l_i}+\sum_{e_j}y^{e_j}+3\sum_{d_k}y^{d_k}+6\sum_{{\mathbf Q}_m}y^{{\mathbf Q}_m} \Big)\\
& ~~~ +2 q_S \Big(2\sum_{l_i}y^{l_i}q_{\mathbf L}^{l_i}\e_{l_i}+\sum_{e_j}y^{e_j}q_{\mathbf L}^{e_j}\e_{e_j}\\
& ~~~~~~~~~~~~ +3\sum_{d_k}y^{d_k}q_{\mathbf L}^{d_k}\e_{d_k}+6\sum_{{\mathbf Q}_m}y^{{\mathbf Q}_m}q_{\mathbf L}^{{\mathbf Q}_m}\e_{{\mathbf Q}_m} \Big)&= -t_{YAA},\\
Tr[q_Aq_Aq_A]_{secluded} &= 
2\sum_{l_i}[(q_{\mathbf L}^{l_i})^3+(q_R^{l_i})^3]+\sum_{e_j}[(q_{\mathbf L}^{e_j})^3+(q_R^{e_j})^3]
\\
& ~~
+3\sum_{d_k}[(q_{\mathbf L}^{d_k})^3+(q_R^{d_k})^3] +6\sum_{{\mathbf Q}_m}[(q_{\mathbf L}^{{\mathbf Q}_m})^3+(q_R^{{\mathbf Q}_m})^3] & \\
& = q_S^3\Big(2\sum_{l_i}\e_{l_i}+\sum_{e_j}\e_{e_j}+3\sum_{d_k}\e_{d_k}+6\sum_{{\mathbf Q}_m}\e_{{\mathbf Q}_m} \Big) & \\
& ~~ -3q_S^2 \Big(2\sum_{l_i}q_{\mathbf L}^{l_i}+\sum_{e_j}q_{\mathbf L}^{e_j}+3\sum_{d_k}q_{\mathbf L}^{d_k}+6\sum_{{\mathbf Q}_m}q_{\mathbf L}^{{\mathbf Q}_m} \Big)\\
& ~~ +3 q_S \Big(2\sum_{l_i}(q_{\mathbf L}^{l_i})^2\e_{l_i}+\sum_{e_j}(q_{\mathbf L}^{e_j})^2\e_{e_j}
\\
& ~~~~~~~~~~~ +3\sum_{d_k}(q_{\mathbf L}^{d_k})^2\e_{d_k}+6\sum_{{\mathbf Q}_m}(q_{\mathbf L}^{{\mathbf Q}_m})^2\e_{{\mathbf Q}_m} \Big)&= -t_{AAA},\\
Tr[q_AT_2T_2]_{secluded} &= 
(\sum_{l_i}\e_{l_i} + 3\sum_{{\mathbf Q}_m}\e_{{\mathbf Q}_m})q_S & = -t_2,\\
Tr[q_AT_3T_3]_{secluded} &= 
(\sum_{d_k}\e_{d_k} + 2\sum_{{\mathbf Q}_m}\e_{{\mathbf Q}_m})q_S & = -t_3.
\end{array}~~~
\eea

\newpage

\bibliography{references}
\bibliographystyle{JHEP}

\end{document}